\documentclass{article}
\pdfoutput=1
\usepackage{jcappub}
\usepackage{graphicx}
\usepackage{color}

\def\lbldef#1#2{\expandafter\gdef\csname #1\endcsname {#2}}

\def\href#1#2{#2}

\newcommand{\ceffs}{c_\mathrm{eff}^2}
\newcommand{\cviss}{c_\mathrm{vis}^2}

\title{Robustness of cosmic neutrino background detection in the cosmic microwave background}
\author[a]{Benjamin Audren,}
\author[b]{Emilio Bellini,}
\author[b]{Antonio J. Cuesta,}
\author[c]{Satya Gontcho A Gontcho,}
\author[d]{Julien Lesgourgues,}
\author[e]{Viviana Niro,}
\author[f,g]{Marcos Pellejero-Ibanez,}
\author[c]{Ignasi P\'erez-R\`afols,}
\author[h]{Vivian Poulin,}
\author[i]{Thomas Tram,}
\author[f,g]{Denis Tramonte}
\author[b,j,k]{and Licia Verde}

\affiliation[a]{Institut de Th\'eorie des Ph\'enom\`enes Physiques, \'Ecole Polytechnique F\'ed\'erale de Lausanne, CH-1015, Lausanne, Switzerland}
\affiliation[b]{Institut de Ci{\`e}ncies del Cosmos, Universitat de Barcelona, IEEC-UB, Mart{\'\i} i Franqu{\`e}s 1, E08028 Barcelona, Spain}
\affiliation[c]{Dept. d'Astronomia i Meteorologia, Institut de Ci{\`e}ncies del Cosmos, Universitat de Barcelona, IEEC-UB, Mart{\'\i} i Franqu{\`e}s 1, E08028 Barcelona, Spain}
\affiliation[d]{CERN, Theory Division, CH-1211 Geneva 23, Switzerland, and LAPTh, Univ. de Savoie, CNRS, B.P.110, Annecy-le-Vieux F-74941, France}
\affiliation[e]{Departamento de F\'isica Te\'orica, Universidad Aut\'onoma de Madrid, and Instituto de F\'isica Te\'orica UAM/CSIC, Calle Nicol\'as Cabrera 13-15, Cantoblanco, E-28049 Madrid, Spain}
\affiliation[f]{Instituto de Astrof\'{\i}sica de Canarias (IAC), C/V\'{\i}a L\'{a}ctea s/n, E-38200, La Laguna, Tenerife, Spain}
\affiliation[g]{Departamento Astrof\'{i}sica, Universidad de La Laguna (ULL), E-38206 La Laguna, Tenerife, Spain}
\affiliation[h]{LAPTh, Univ. de Savoie, CNRS, B.P.110, Annecy-le-Vieux F-74941, France}
\affiliation[i]{Institute of Cosmology and Gravitation, University of Portsmouth, Dennis Sciama Building, Burnaby Road, Portsmouth PO1 3FX, United Kingdom}
\affiliation[j]{ICREA (Instituci\'o catalana de recerca i estudis avan\c{c}ats)}
\affiliation[k]{Institute of Theoretical Astrophysics, University of Oslo, 0315 Oslo, Norway}

\abstract{The existence of a cosmic neutrino background can be probed indirectly by CMB experiments, not only by measuring the background density of radiation in the universe, but also by
searching for the typical signatures of the fluctuations of free-streaming species in the temperature and polarisation power spectrum. Previous studies have already proposed a rather
generic parametrisation of these fluctuations, that could help to discriminate between the signature of ordinary free-streaming neutrinos, or of more exotic dark radiation models. Current data are compatible with standard values of these parameters, which seems to bring further evidence for the existence of a cosmic neutrino background. In this work, we investigate the robustness of 
this conclusion under various assumptions. We generalise the definition of an effective sound speed and viscosity speed to the case of massive neutrinos or other dark radiation components experiencing a non-relativistic transition. We show that current bounds on these effective parameters do not vary significantly when considering an arbitrary value of the particle mass, or extended cosmological models with a free effective neutrino number, dynamical dark energy or a running of the primordial spectrum tilt. We conclude that it is possible to make a robust statement about the detection of the cosmic neutrino background by CMB experiments.}

\begin{document}

\hfill {\small CERN-PH-TH-2014-266, LAPTH-238/14, }\\
\vglue -0.3cm
\hfill {\small FTUAM-14-51, IFT-UAM/CSIC-14-132} \\[3mm]

\maketitle
\hypersetup{pageanchor=true}

\section{Introduction}

Neutrinos are the only dark matter component that has been directly detected.  Despite neutrinos not being cold and not being the bulk of the dark matter in the Universe, they are a particularly interesting component to study. Not only because of the synergy between astrophysical observations and particle physics experiments, but also because they contribute a large fraction of the energy density in the Universe during the radiation dominated stage.
The first indirect confirmation of the existence of a cosmological neutrino background has been obtained by assuming standard neutrino properties, and adding only one extra parameter to the standard $\Lambda$CDM model: the effective number of neutrino species, $N_{\rm eff}$, equal to $3.046$\footnote{The number of  (active) neutrinos species is 3. As
the neutrino decoupling epoch was immediately followed by $e^+e^-$ annihilation, the value of $N_{\rm eff}$ for 3 neutrino species is slightly larger than 3.} \cite{Mangano:2005cc} in the standard model. By using Cosmic Microwave Background (CMB) observations, the WMAP collaboration showed to high statistical significance that $N_{\rm eff}>0$ \cite{Dunkley2008WMAP, Komatsu2008WMAP}, yielding therefore a confirmation, albeit indirect, of the existence of the cosmic neutrino background.
With recent data from Planck, $N_{\rm eff}=0$ is disfavoured at the level of about 10$\sigma$ \cite{PlanckXVI}.

But $N_{\rm eff}$ does not only count the number of neutrino species. Even assuming standard neutrino physics, departures from $N_{\rm eff}$ could be caused by any ingredient contributing to the expansion rate of the Universe in the same way as a radiation background. The possibilities for this extra ingredient are many: extra relativistic particles (either decoupled, self-interacting, or interacting with a dark sector), a background of gravitational waves, an oscillating scalar field with quartic potential, departures from Einstein gravity, large extra dimensions or something else. Such a component is usually dubbed ``dark radiation" \cite[e.g.,][]{Ackermanetal, Abazajian:2012ys,Weinberg:2013kea,Kelso:2013paa, Mastache:2013iha,DiBari:2013dna, Archidiacono:2013fha, Boehm:2012gr, Hasenkamp:2012ii, Anchordoqui:2014dpa, SolagurenBeascoa:2012cz,GonzalezGarcia:2012yq}.
In principle, we could even assume that the cosmic neutrino background does not exist, while another dark radiation component explains the measured value of $N_{\rm eff}$.

It is well known that free streaming particles like decoupled neutrinos leave specific signatures  on the CMB, not only through their contribution to the background evolution, but also because their density/pressure perturbations, bulk velocity and anisotropic stress are additional sources for the gravitational potential via the Einstein equations (see  for example \cite{Bashinsky:2003tk,Hou:2011ec,Lesgourgues:1519137}  and references therein for a detailed discussion). On that basis, several analyses have shown that the CMB can make a  more precise statement on the existence of a cosmic neutrino background in the Universe than by just measuring $N_\mathrm{eff}>0$ and showing that it is compatible with  the standard value.
The CMB seems to prove that the perturbation of neutrinos -- or more precisely, the perturbation of free-streaming particles with the required abundance -- are needed to explain the data.

The strategy of several recent papers \cite{Archidiacono:2011gq,Archidiacono:2012gv,Diamanti:2012tg,Archidiacono:2013lva, Gerbinoetal} was to introduce\footnote{Indeed we are referring here to the definition of ($\ceffs$, $\cviss$) first introduced by these authors. This parametrisation is however strongly inspired from earlier works, e.g., \cite{Hu:1998tk,Hu:1998GDM,Bashinsky:2003tk,Trotta:2004ty,Smith:2011es}.}  two phenomenological parameters, $c_\mathrm{eff}$ and $c_\mathrm{vis}$. The effect of the parameter $\ceffs$ is to generalize the linear relation between isotropic pressure perturbations and density perturbations, while $\cviss$ directly modifies the anisotropic stress equation for neutrinos. These parameters allow to distinguish the perturbations of relativistic free-streaming species, corresponding to ($\ceffs$, $\cviss$) = (1/3,~1/3), from those of a perfect relativistic fluid with ($\ceffs$, $\cviss$) = (1/3,~0), or a scalar field scaling like radiation with ($\ceffs$, $\cviss$) = (1,~0), or a more general case with arbitrary ($\ceffs$, $\cviss$). Self-interacting neutrinos or other types of dark radiation candidates might not be exactly equivalent to these models with definite and constant value of ($\ceffs$,$\cviss$) (see for instance~\cite{Cyr-Racine:2013jua,Oldengott:2014qra}), but this parametrisation is considered flexible enough for providing a good approximation to several alternatives to the standard case of free-streaming particles. We will come back to the motivations for this parametrisation in section 2.

Previous works found that the allowed window for $\ceffs$ is shrinking close to $1/3$, and that the data starts to be very sensitive also to $\cviss$, although this parameter has a smaller effect.
For instance, using Planck 2013 data, ref.~\cite{Gerbinoetal} obtained $(\ceffs, \cviss) = (0.304\pm0.026, 0.60\pm0.36)$ at the 95\% CL. The next Planck data release is expected to bring even better sensitivity, thanks to better temperature and new polarisation data.

However, recent results on ($\ceffs$, $\cviss$) were derived in the context of the minimal $\Lambda$CDM model, with negligible neutrino masses. The point of the present paper is to answer the two important questions: Are these bounds stable when considering massive neutrinos, instead of the purely massless limit? And could ($\ceffs$, $\cviss$) be degenerate with other cosmological parameters, like e.g., $N_\mathrm{eff}$, a running of the primordial spectrum index, or the equation of state of dynamical dark energy? These issues are important to better assess the meaning of current bounds, and also to prepare the interpretation of future results. Indeed, if future data bring stronger evidence for standard neutrino perturbations, we will need to understand whether such conclusions are robust or model-dependent. On the other hand, if a deviation from the standard behaviour is found in the context of the minimal $\Lambda$CDM model, we will need to know whether extended cosmological models have the potential to reconcile observations with standard values of ($\ceffs$, $\cviss$).
The rest of this paper is organised as follows: In section~2 we  present the set of equations describing a massless relativistic component with arbitrary $(\ceffs, \cviss)$, and its generalisation to the case of species becoming non-relativistic at late times. In section~3 we analyse the physical  effect of the phenomenological parameters on the observables. In section~4 we describe our methodology and introduce the data sets used. We present our results in section~5 and we discuss and conclude in section~6.

\section{Modelling the properties of the (dark) radiation component \label{sec:model}}

While the parameter $N_{\rm eff}$  affects  the expansion rate of the early universe,   we want to introduce  some parameters describing the behaviour of perturbations. If we were comparing ordinary neutrinos with a concrete physical model (e.g., neutrinos with a given collision or self-interaction term \cite{Wilkinson:2014ksa,Archidiacono:2014nda}, oscillating scalar field with quartic potential, etc.), there would be no ambiguity in the set of equations and parameters to compare with data. We are not in this situation: we want to define some effective parameters, chosen to provide an exact or approximate description of a wide variety of non-standard models for the radiation component in the universe. From now on,  we follow the notations of Ma \& Bertschinger \cite{Ma:1995ey}. 

The logic followed by previous authors and leading to the definition of ($\ceffs$, $\cviss$) is to postulate a linear relation between isotropic pressure perturbations and density perturbations given by a squared sound speed $\ceffs$, assumed for simplicity to be independent of time. The approach is then extended to anisotropic pressure by introducing another constant, the viscosity coefficient $\cviss$.

Technically, this amounts in writing the usual continuity and Euler equations, valid for any decoupled species, and replacing the pressure perturbation $\hat{\delta p}$ by $\ceffs \hat{\delta \rho}$. The hats mean that we are referring to the pressure and density defined in the frame (or in the gauge) comoving with the fluid we are studying, i.e., in which the energy flux divergence $\theta$ vanishes. From the gauge transformations~\cite{Ma:1995ey} one can show that in an arbitrary gauge, the density perturbations $\delta \rho$, the pressure perturbation $\delta p$ and the energy flux divergence $\theta$ are related to the comoving density/pressure perturbations by
\begin{eqnarray}
\hat{\delta \rho} &=& \delta \rho + 3 \frac{\dot{a}}{a} (1 + w_{\rm dr}) \bar{\rho} \frac{\theta}{k^2} \\
\hat{\delta p} &=& \delta p + 3 \frac{\dot{a}}{a} (1+w_{\rm dr}) c_a^2  \bar{\rho} \frac{\theta}{k^2}  \label{comoving_pressure}
\end{eqnarray}
where $a$ is the usual scale factor, the dot indicates derivative with respect to conformal time, $w_{\rm dr} \equiv \bar{p}/\bar{\rho}$ and $c_a^2 \equiv \dot{\bar{p}}/\dot{\bar{\rho}}$. The pressure perturbation appears as a source term in the continuity equation and the Euler equation (see eq.~(29, 30) of \cite{Ma:1995ey}). If we assume $\hat{\delta p} = \ceffs \hat{\delta \rho}$, we should replace $\delta p$ in these two places by
\begin{equation}
\delta p = \ceffs \left(\delta \rho + 3 \frac{\dot{a}}{a} (1 + w_{\rm dr}) \bar{\rho} \frac{\theta}{k^2} \right) - 3 \frac{\dot{a}}{a} (1+w_{\rm dr}) c_a^2  \bar{\rho} \frac{\theta}{k^2} \,.\label{deltap_general}
\end{equation}

\subsection{Massless neutrinos}

In the relativistic limit, eq.~(\ref{deltap_general}) becomes
\begin{equation}
\frac{\delta p}{\bar{\rho}} = \ceffs \left(\delta + 4 \frac{\dot{a}}{a}  \frac{\theta}{k^2} \right) - \frac{4}{3} \frac{\dot{a}}{a}  \frac{\theta}{k^2} \,.\label{deltap_rel}
\end{equation}
For decoupled massless neutrinos, the Boltzmann equation can be integrated over momentum, leading to a Boltzmann hierarchy in which the first two equations are equivalent to the continuity and Euler equation. Replacing the two occurrences of $\delta p$ in these equations by the above expression gives:
\begin{eqnarray}
\dot{\delta}_\nu &=&
\left( 1 - 3 \ceffs \right) \frac{\dot{a}}{a}
\left(\delta_\nu+\frac{4}{k^2} \frac{\dot{a}}{a} \theta_\nu \right)
-\frac{4}{3} (\theta_\nu +M_\mathrm{continuity})\,,
\\
\dot{\theta}_\nu &=&
\frac{k^2}{4} (3\ceffs) \left(\delta_\nu+\frac{4}{k^2} \frac{\dot{a}}{a} \theta_\nu\right)
- \frac{\dot{a}}{a} \theta_\nu - k^2 \sigma_\nu + M_\mathrm{Euler}\,,
\end{eqnarray}
where the subscript $\nu$ refers to the neutrino (or dark radiation) component. The above equations are valid in any gauge provided that the two quantities ($M_\mathrm{continuity}$, $M_\mathrm{Euler}$) refer to the right combination of metric perturbations, e.g. $( \dot{h}/2,0)$ in the synchronous gauge and $(-3 \dot{\phi}, k^2 \psi)$ in the Newtonian gauge (see \cite{Ma:1995ey} for the definition of $h$, $\phi$ and $\psi$).
When $\ceffs$ is set to $1/3$, the standard equations are recovered, since for relativistic free-streaming species the sound speed squared is exactly $1/3$.

While $\delta p$ appears a source term for $\delta$ and $\theta$, the anisotropic pressure $\sigma$ is sourced
in the next equation of the Boltzmann hierarchy by $\theta+M_\mathrm{shear}$. Extending the previous logic to the level of anisotropic pressure can be done by multiplying this source term by $(3 \cviss)$. Then, for $\cviss=1/3$, standard equations will be recovered by construction. This prescription leads to:
\begin{equation}
\dot{F}_{\nu2} = 2 \dot{\sigma}_\nu = (3  \cviss) \frac{8}{15} (\theta_\nu +  M_\mathrm{shear}) - \frac{3}{5} k F_{\nu 3}\,, \label{sigdot_relativistic}
\end{equation}
where $F_{\nu \ell}$ are the Legendre multipoles of the momentum integrated neutrino distribution function as defined in ref.~\cite{Ma:1995ey}.
$M_\mathrm{shear}$ is $0$ in the Newtonian gauge and given by $(\dot{h}+6\dot{\eta})/2$ in the synchronous gauge.

The next equations in the hierarchy are left unmodified.
A coefficient $\cviss$ was first introduced by Hu \cite{Hu:1998GDM}, as an approximate way to close the Boltzmann hierarchy at order $l=2$. For that purpose, the term $F_{\nu 3} $ was eliminated from equation (\ref{sigdot_relativistic}). The above parametrisation was introduced later in ref.~\cite{Archidiacono:2011gq}, keeping that term, in order to recover the standard equations in the limit  $\cviss=1/3$. The limit $\cviss=0$ describes a species with isotropic pressure (like, for instance, a perfect fluid), since in that limit, ${\sigma}_\nu$ and all multipoles $F_{\nu \ell}$ with $\ell \geq 3$ remain zero at all times.

\subsection{Massive neutrinos}

We will now present original results, showing how the previous parametrisation can be extended to the case of light relics experiencing a non-relativistic transition such as massive neutrinos. In the massive neutrino case, the Boltzmann equation cannot be integrated over momentum, and one must solve one hierarchy per momentum bin. We wish to introduce the ($\ceffs$, $\cviss$) factors in the same way as for massless neutrinos, assuming for simplicity that they affect each momentum equally. The strategy is again to identify the source terms corresponding to $\hat{\delta p}$ in the continuity/Euler equation and multiply them by $(3 \ceffs)$, and similarly to identify the source term for $\sigma$ in the quadrupole equation and multiply it by $(3 \cviss)$.

One can define several statistical momenta of the background phase-space distribution $f_0(q)$, including the usual background density $\bar{\rho}$ and pressure $\bar{p}$, and also a quantity called the pseudo-pressure in \cite{Shoji:2010hm}:
\begin{equation}
\tilde{p} = \frac{4\pi}{3} a^{-4} \int_0^\infty dq \frac{q^6}{\epsilon^3} f_0(q)~,
\end{equation}
where $\epsilon$  is the comoving energy of the particle. Throughout this paper, we use the Boltzmann code {\sc class}\footnote{Code available at \url{http://class-code.net} and \url{https://github.com/lesgourg/class_public}.}~\cite{Lesgourgues:2011re,Blas:2011rf} to compute observable spectra. It happens that the pseudo-pressure is always computed by {\sc class}, because it enters into the expression of the fluid approximation switched on deep inside the Hubble radius \cite{Lesgourgues:2011rh}. Pseudo-pressure is also useful in the present context, since the comoving pressure perturbation $\hat{\delta p}$ of eq.~(\ref{comoving_pressure}) can also be expressed as
\begin{equation}
\hat{\delta p} = \delta p + \frac{\dot{a}}{a} (5 \bar{p} - \tilde{p}) \frac{\theta}{k^2}~.
\end{equation}
One can write down the continuity and Euler equation, decomposing each perturbation as an integral over momentum, involving the Legendre momenta of the perturbed phase-space distribution $\Psi_l(k, \tau, q)$. Then, like for massless neutrinos, we identify the two terms involving $\hat{\delta p}$ and replace them by
\begin{equation}
\delta p = \ceffs \left(\delta \rho + 3 \frac{\dot{a}}{a} (\bar{\rho} + \bar{p}) \frac{\theta}{k^2} \right) - \frac{\dot{a}}{a}
(5 \bar{p} - \tilde{p}) \frac{\theta}{k^2} ~.
\end{equation}
Finally, assuming that $\ceffs$ is a momentum-independent coefficient\footnote{We shall discuss this assumption \textit{a posteriori} in the Conclusions}, we can remove the integral over $q$ and obtain a modified Boltzmann hierarchy for each momentum $q$:
\begin{eqnarray}
\dot{\Psi}_0
&=& \frac{\dot{a}}{a} \left( 1- 3 c_\mathrm{eff}^2 \right) \frac{q^2}{\epsilon^2} \left[  \Psi_0
+ 3 \frac{\dot{a}}{a}  \frac{5p-\tilde{p}}{\rho+p} \frac{\epsilon }{k q} \Psi_1 \right]
- \frac{qk}{\epsilon} \Psi_1 + \frac{1}{3} M_\mathrm{continuity} \frac{d \ln f_0}{d \ln q}~,\\
\dot{\Psi}_1 &=& c_\mathrm{eff}^2 \frac{q k}{\epsilon} \left[ \Psi_0 + 3 \frac{\dot{a}}{a} \frac{5p-\tilde{p}}{\rho+p} \frac{\epsilon}{qk} \Psi_1 \right] - \frac{\dot{a}}{a} \frac{5p-\tilde{p}}{\rho+p} \Psi_1 - \frac{2}{3} \frac{qk}{\epsilon} \Psi_2
- \frac{\epsilon}{3qk} M_\mathrm{euler} \frac{d \ln f_0}{d \ln q}~.
\end{eqnarray}
Finally, in the $l=2$ equation, we multiply again the source term of the shear by $(3 \cviss)$ and obtain:
\begin{equation}
\dot{\Psi}_2 = \frac{q k}{5 \epsilon} \left(6 c_\mathrm{vis}^2 \Psi_1 - 3 \Psi_3 \right) - 3 c_\mathrm{vis}^2 \frac{2}{15} M_\mathrm{shear}  \frac{d \ln f_0}{d \ln q} \,. \\
\end{equation}
Higher momenta in the Boltzmann hierarchy are left unchanged. Again, when $(\ceffs, \cviss) = (1/3, 1/3)$, we recover exactly standard equations.

\section{Impact of $(\ceffs, \cviss)$ on observables}

We implemented the previous equations of motion into {\sc class} in order to study the impact of $(\ceffs, \cviss)$  on observable quantities. There is no need to modify initial conditions, because on super-Hubble scales perturbations are insensitive to pressure gradients, and hence to $\ceffs$. The perturbations also have negligible anisotropic pressure in the super-Hubble limit, so $\cviss$ is not playing a role either. Unless otherwise stated, for all parameters that take fixed values, we adopt the same settings as in the ``base model'' of the Planck 2013 parameter paper~\cite{PlanckXVI}.

\subsection{Effect on neutrino perturbations}

\begin{figure}
\includegraphics[width=0.5\textwidth]{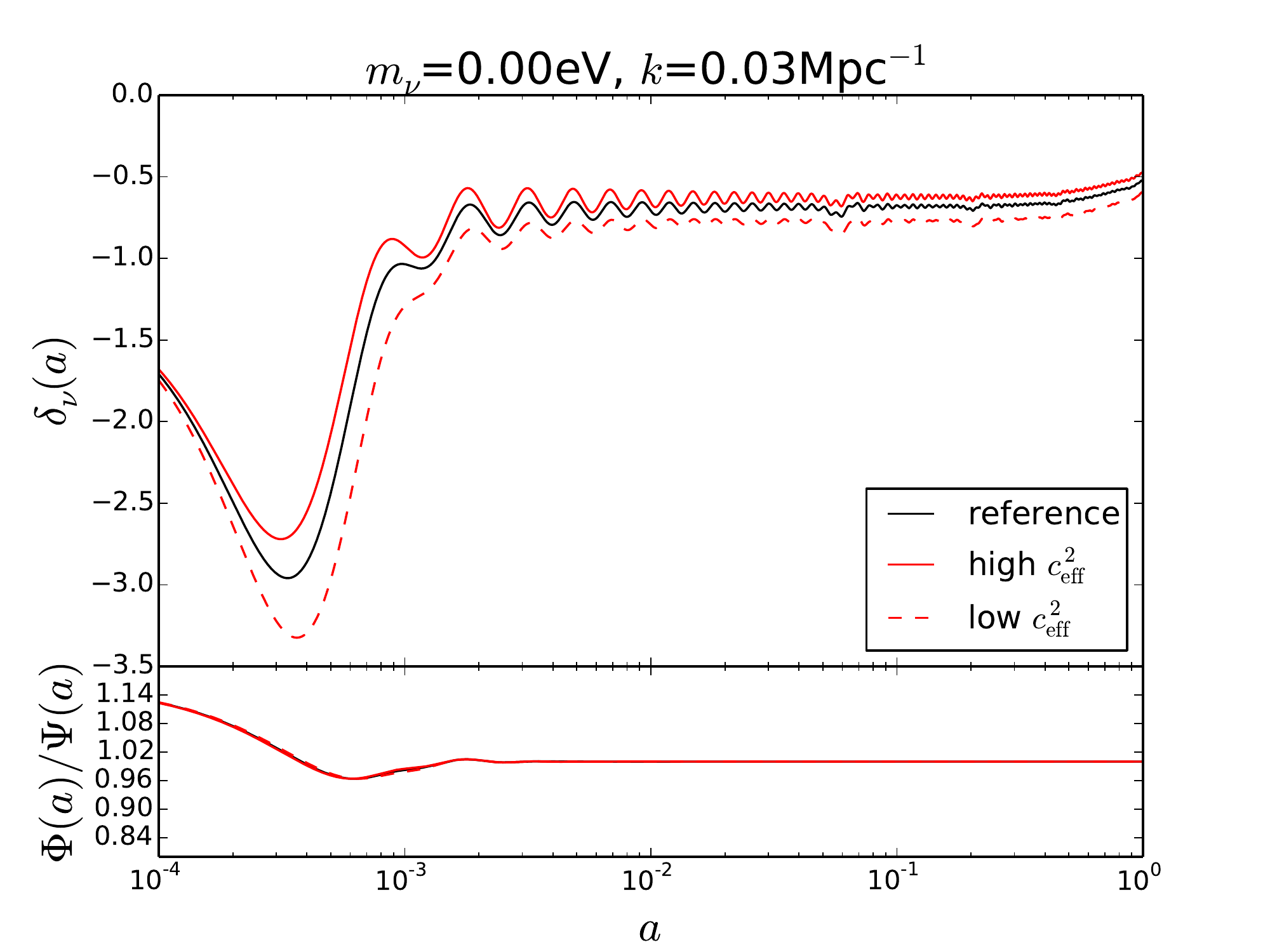}
\includegraphics[width=0.5\textwidth]{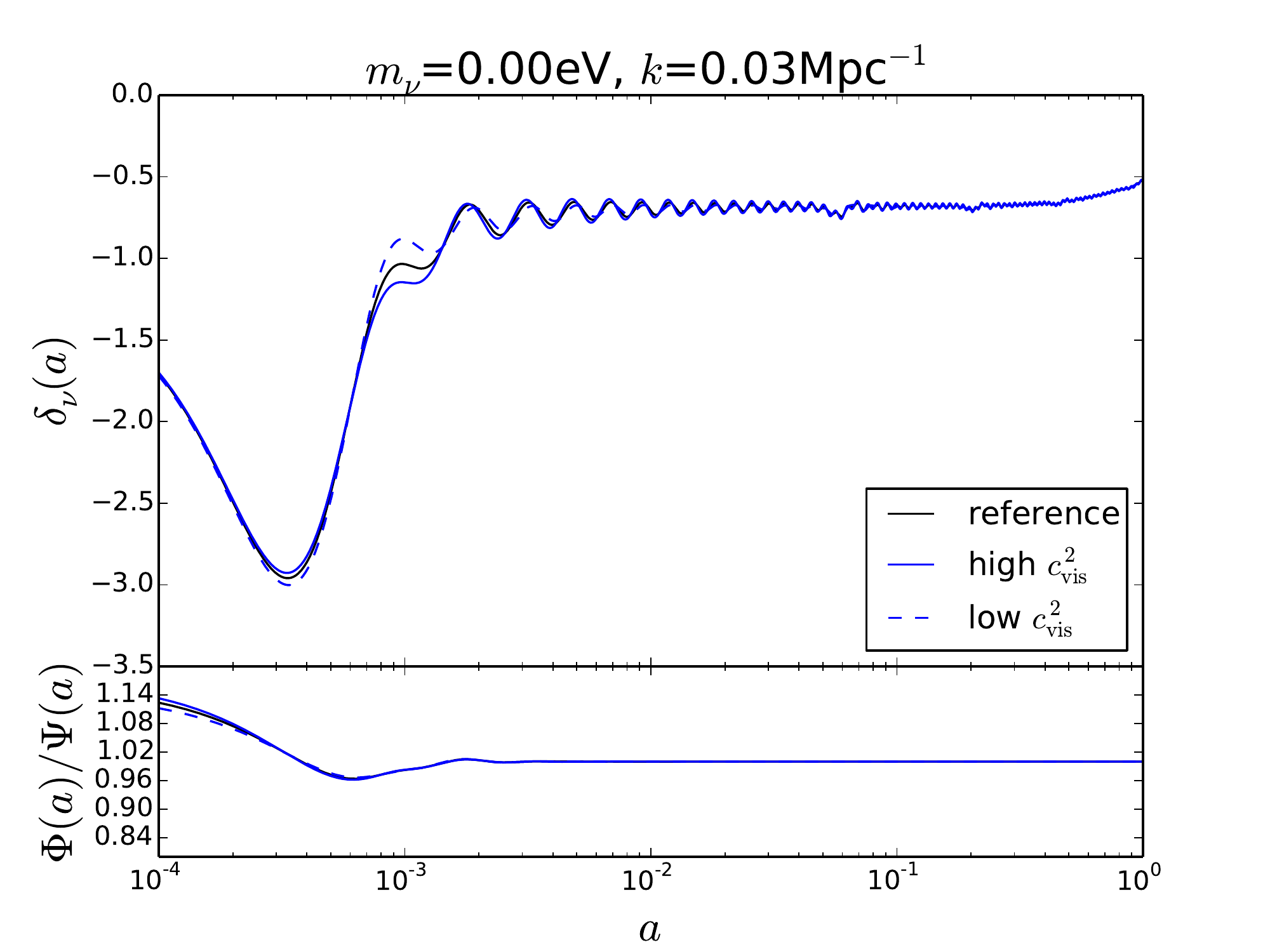}
\\
\includegraphics[width=0.5\textwidth]{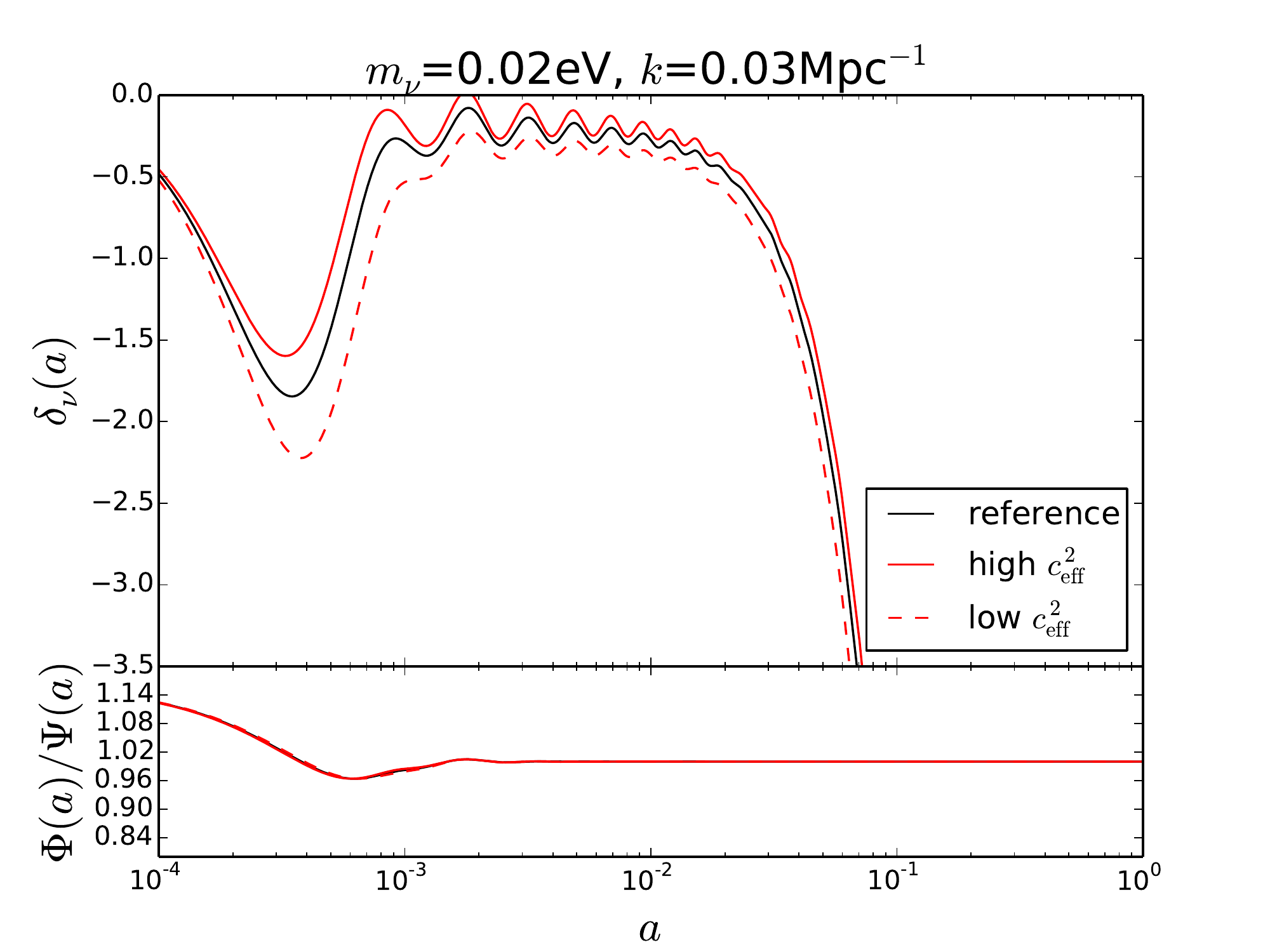}
\includegraphics[width=0.5\textwidth]{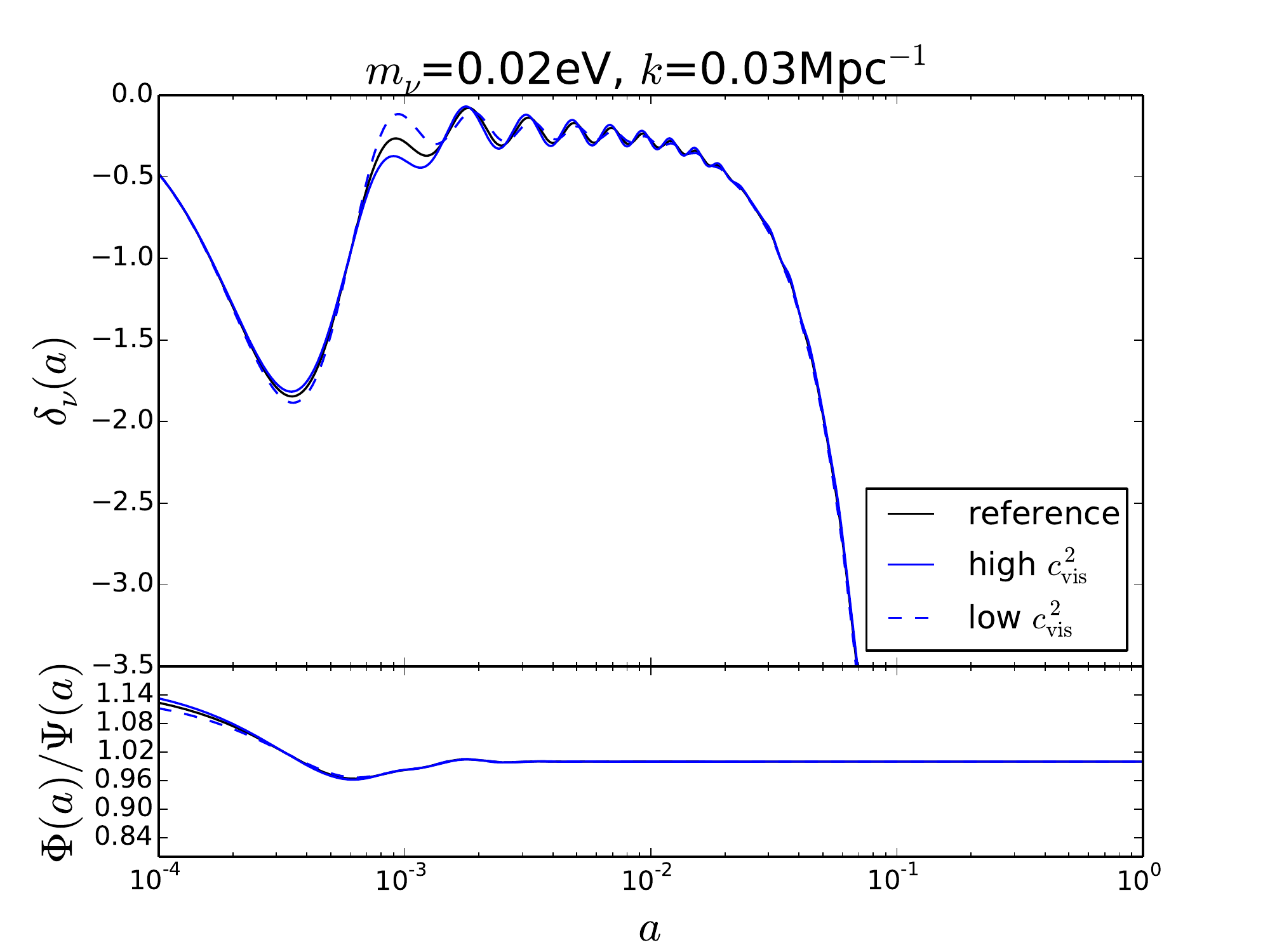}
\\
\includegraphics[width=0.5\textwidth]{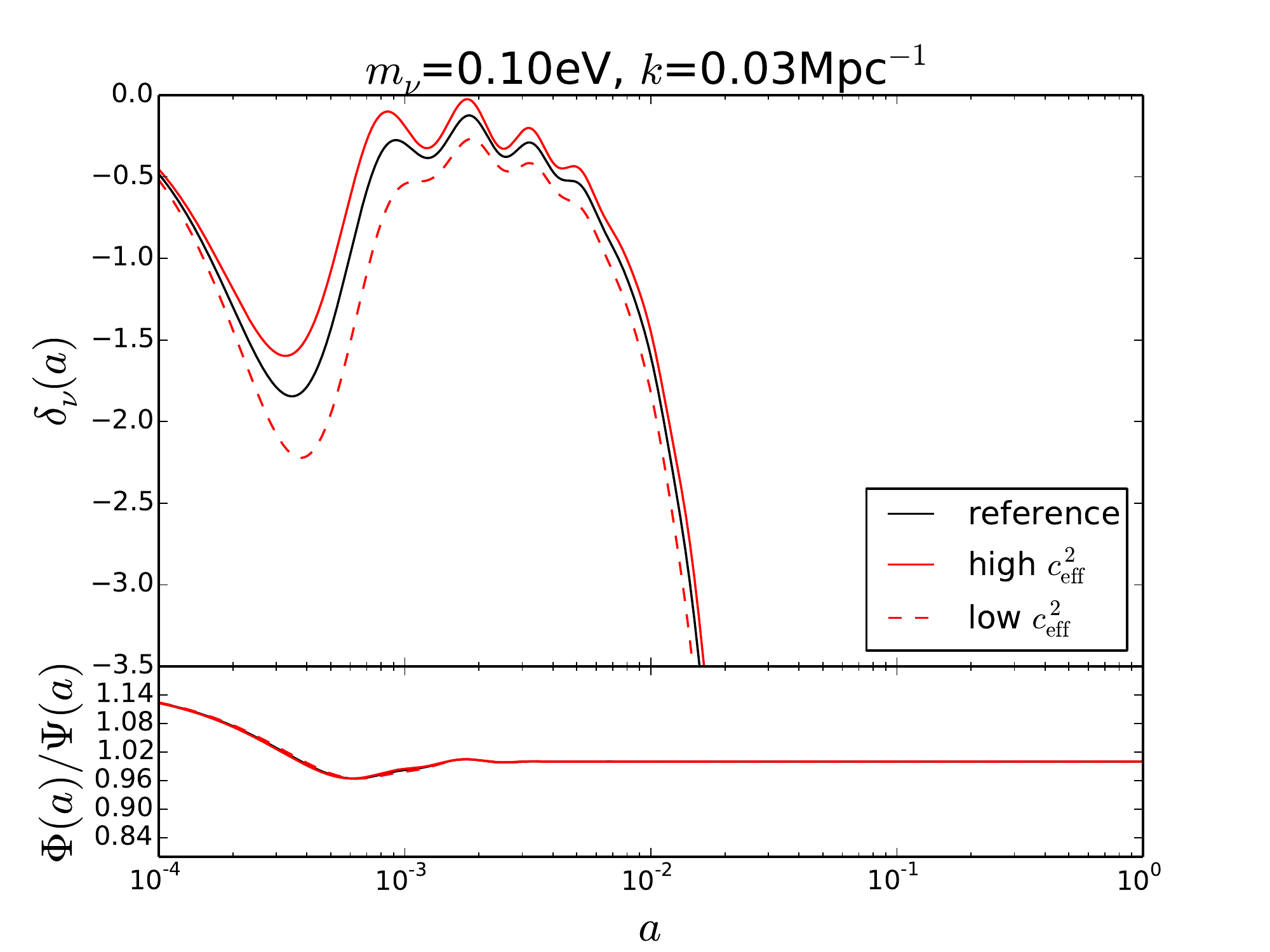}
\includegraphics[width=0.5\textwidth]{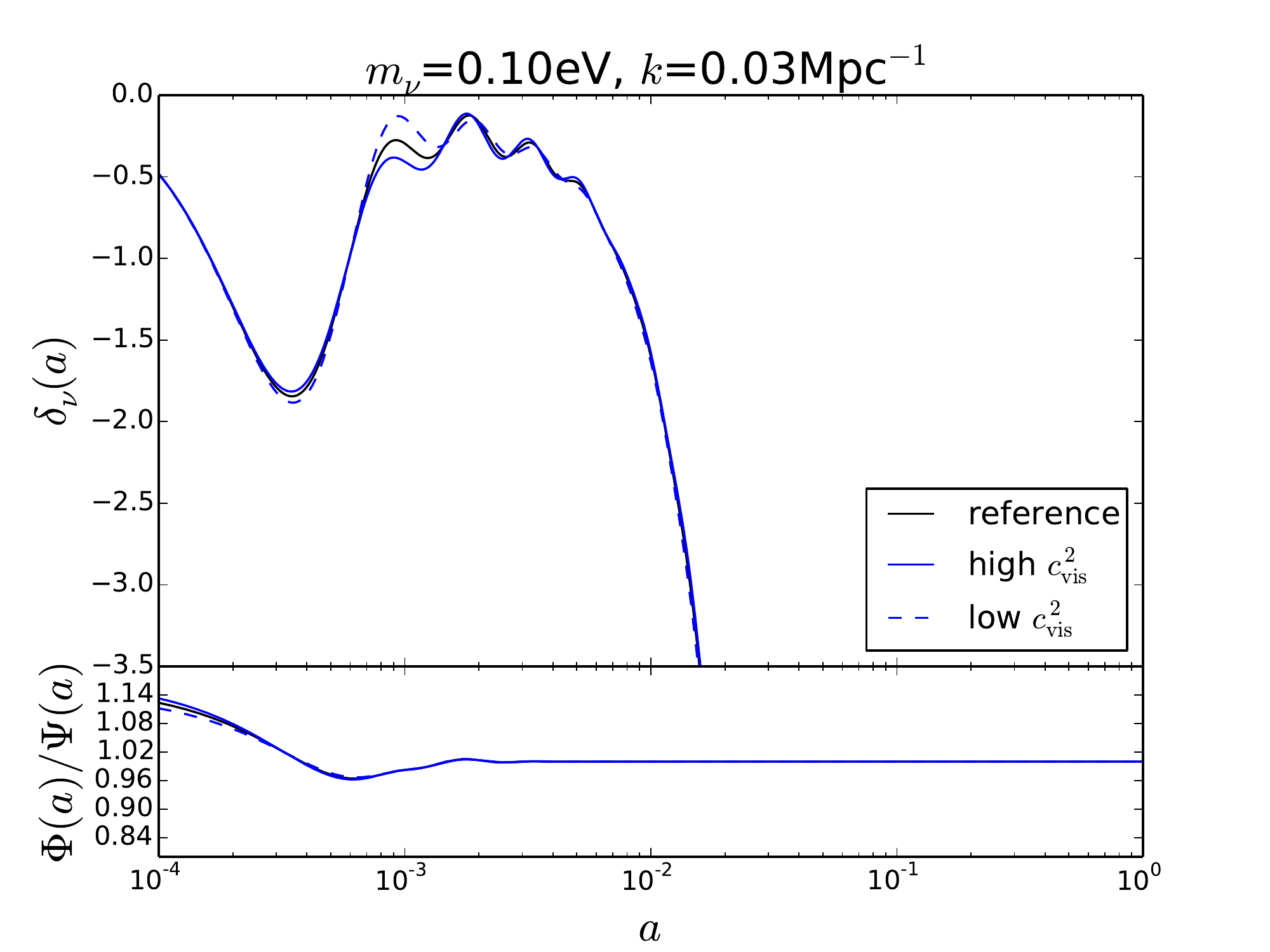}
\caption{Neutrino density perturbations as a function of scale factor for a $\Lambda$CDM model with massless neutrinos (top panels), three degenerate neutrinos with $m_{\nu}=0.02$~eV each (middle panels), 
and $m_{\nu}=0.10$~eV (bottom panels). All panels show the evolution of the perturbations for a fixed scale of 0.03~Mpc$^{-1}$. Solid black lines show a reference model with $c^2_{\rm eff}=c^2_{\rm vis}=1/3$. In the left panels, solid red lines and dashed red lines correspond to $c^2_{\rm eff}=0.36$ and $0.30$ respectively, whereas in the right panels solid blue lines and dashed blue lines correspond to $c^2_{\rm vis} = 0.36$ and $0.30$ respectively. For reference, the evolution of the ratios of the gravitational potentials are shown for every case.}\label{FIG:deltaUR}
\end{figure}

In figure\ \ref{FIG:deltaUR} we plot the time evolution of the neutrino density perturbations ($\delta_\nu$) and the ratio of the metric fluctuations\footnote{$\Phi$ and $\Psi$ are two gauge-independent combinations of scalar metric fluctuations, equivalent to the Bardeen potentials up to minus signs, and coinciding in the Newtonian gauge with the metric fluctuations $\phi$ and $\psi$ such that $ds^2=-(1+2\psi)dt^2 + a^2 (1-2\phi) d\vec{x}^2$.}   ($\eta\equiv\Phi/\Psi$) at a fixed scale $k=0.03~\textrm{Mpc}^{-1}$. We show the case of (three) massless neutrinos (top panels) and the case of  (three degenerate)  massive neutrinos with $m=0.02$~eV per species (middle panels) and 0.1~eV per species (bottom panels). We have chosen five models in these plots, one reference model in which $c^2_{\rm eff}=c^2_{\rm vis}=1/3$, two models in which we set $c^2_{\rm eff}$ to 0.30 and 0.36, and two models that correspond to $c^2_{\rm vis}$ set to 0.30 and 0.36. Note that on these plots $\delta_\nu$ is always negative: this is because we choose a mode normalised arbitrarily to positive curvature perturbation (i.e., positive gravitational potential) at initial time.

In general, after entering the Hubble radius, the perturbations of a given component grow as a power law of the scale factor ($\delta\propto a^{1+3 w}$) above the sound-horizon (hereafter SH), and start oscillating with a decaying amplitude below the SH. The effective SH of a particular species is defined as
$$
s_\textrm{eff}=\int c_\textrm{eff} d\tau= c_\textrm{eff} \tau\,,
$$
where $\tau$ is conformal time, and the last equality holds for constant sound speeds. Therefore, it is clear that increasing the squared sound speed $\ceffs$, the time at which perturbations stop growing by entering the SH decreases. We expect then a bigger amplitude of the density perturbations $|\delta_\nu|$ for models with lower values of $\ceffs$. Inside the SH, fluctuations are damped, with an oscillatory pattern  $\sim \cos (k c_\textrm{eff} \tau)$ depending on the SH and hence on the effective sound speed. But they are not completely erased: they reach an equilibrium value depending precisely on the pressure to density perturbation ratio. Models with a smaller $\ceffs$ have less pressure perturbations, and hence keep a higher residual density contrast $|\delta_\nu|$ at equilibrium. The decrease of the density contrast observed at late times for massless neutrinos (upper panels) is due to cosmological constant domination ($\Lambda$ suppresses density perturbations by diluting them with the accelerated expansion). Finally, when neutrinos become non-relativistic, their pressure perturbation becomes negligible and they start to collapse gravitationally. A smaller value of $\ceffs$ implies that the pressure perturbation becomes negligible a bit earlier, so the density contrast $|\delta_\nu|$ grows earlier, and moreover starting from a larger equilibrium value. In summary, a smaller $\ceffs$ implies a larger density contrast $|\delta_\nu|$ at all times between the approach of SH crossing and today, and this is what we observe on the left panels of figure\ \ref{FIG:deltaUR}.

The viscosity parameter $c_\textrm{vis}$ mimics the effect of increasing or decreasing the mean free path of particles in an imperfect fluid with interactions. The limit $c_\textrm{vis}=0$ corresponds to a negligible mean free path, i.e., to the strongly interacting regime where the pressure remains isotropic.
A small decrease of $\cviss$ below 1/3 implies that it takes more time for neutrinos to transfer power from a monopole and dipole pattern (i.e., density and velocity perturbations) to a quadrupole pattern (i.e., anisotropic pressure/stress $\sigma_\nu$), like in a weakly interacting fluid with less viscosity. Once the quadrupole is excited, power is transferred to even higher multiples and density fluctuations are damped. Hence the main effect of $c_\textrm{vis}$ is to change slightly the evolution of $\delta_{\nu}$ near the SH crossing time, which is precisely the time at which the anisotropic stress is excited. Models with a smaller $\cviss$ keep a larger density contrast for a slightly longer time. Then the density reaches the damped oscillation regime in slightly more or less time, so the phase of the oscillations is slightly affected by $\cviss$.

In the lower part of each plot, we can see that at early times the evolution of the ratio of the two gravitational potentials $\Phi$ and $\Psi$ is weakly model dependent. In particular, by varying the viscosity parameter, we change the offset between the two metric fluctuations, controlled by the traceless transverse Einstein equation
\begin{equation}
k^2(\Phi-\Psi) = 12 \pi G a^2 (\rho+p) \sigma~.
\end{equation}
The total anisotropic stress on the right-hand side receives contribution from photon perturbations after photon decoupling, and also from neutrino perturbations until their power is transferred to higher multipoles after SH crossing. In models with a lower  $\cviss$, the neutrino anisotropic stress grows more slowly before SH crossing, leading to a reduced difference between the two potentials.

\subsection{CMB temperature and polarisation}

\begin{figure}
\includegraphics[width=0.5\textwidth]{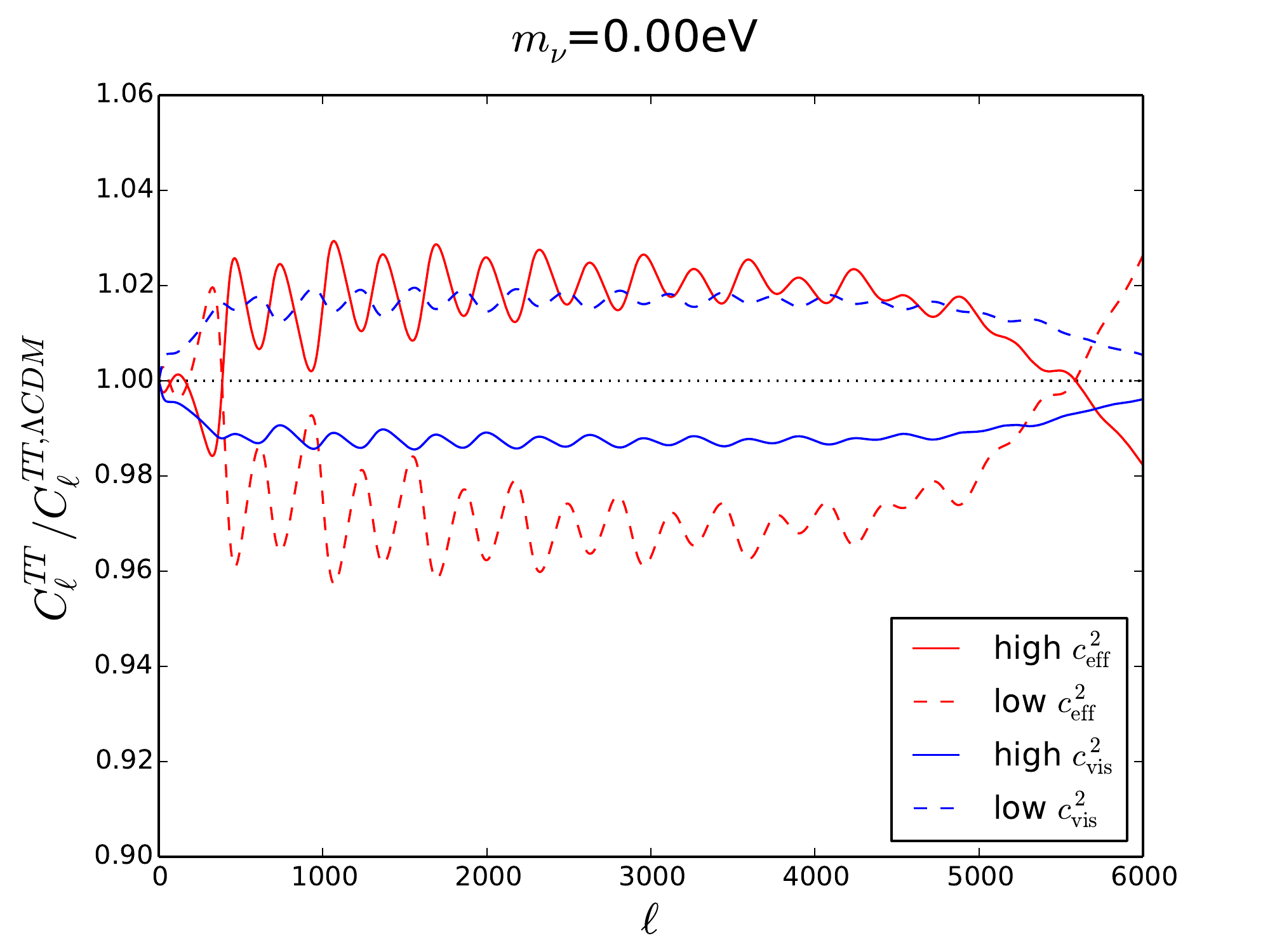}
\includegraphics[width=0.5\textwidth]{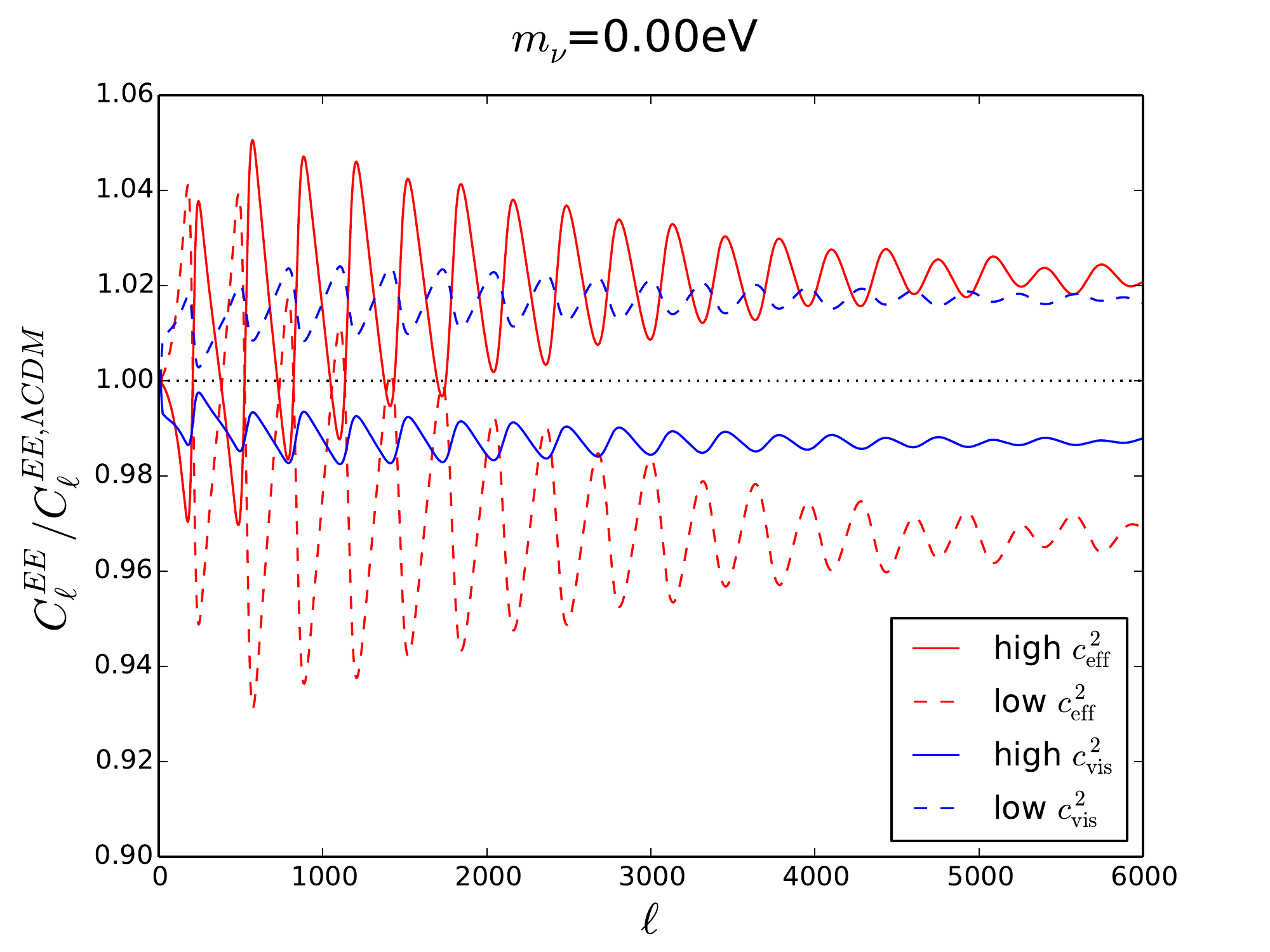}
\\
\includegraphics[width=0.5\textwidth]{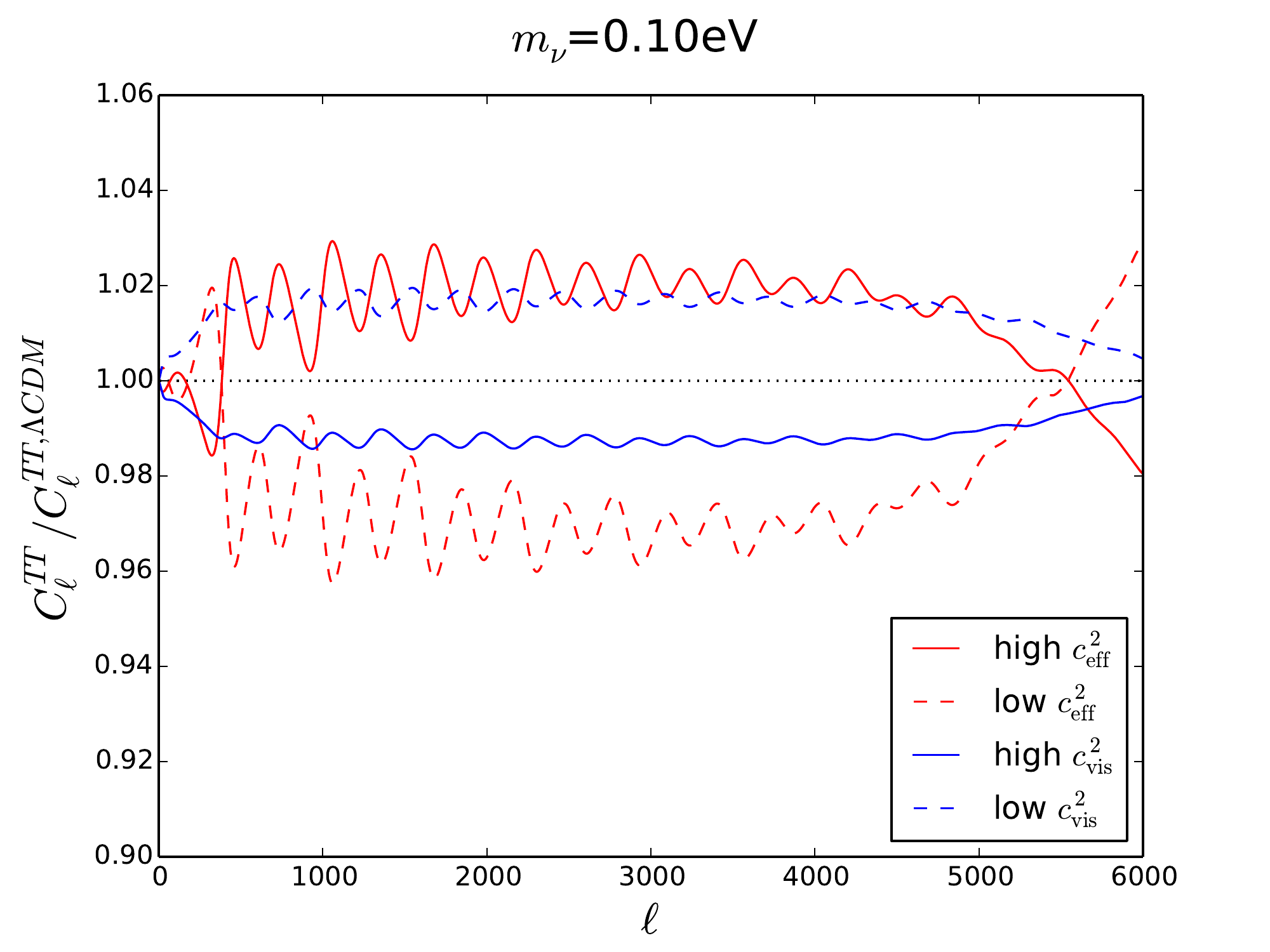}
\includegraphics[width=0.5\textwidth]{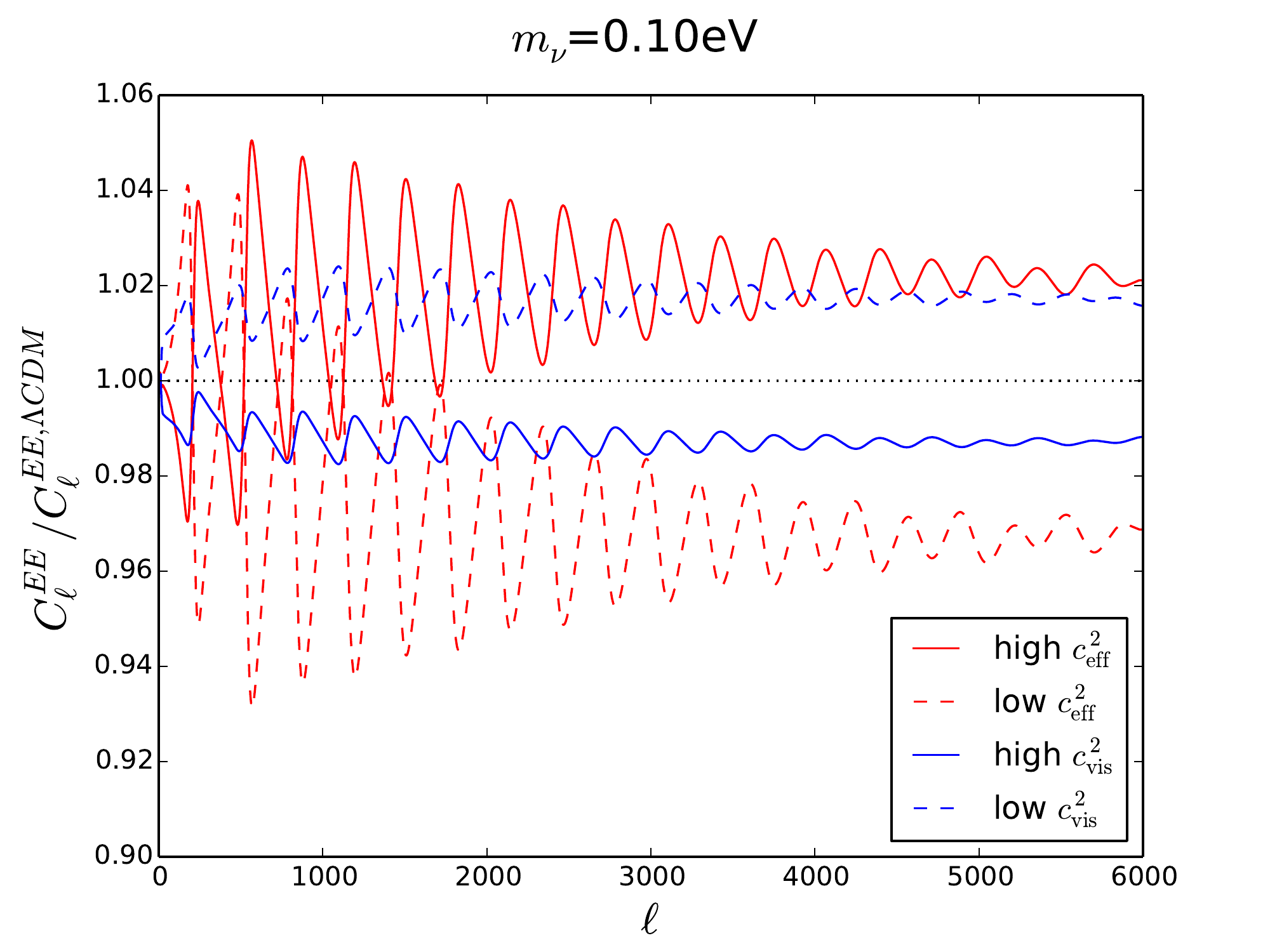}
\caption{CMB power spectrum multipoles for the temperature (left column) and $E$-mode polarisation (right column) for a $\Lambda$CDM model with massless neutrinos (top panels), and three degenerate neutrinos with $m_{\nu}=0.10$~eV (bottom panels). All models are normalised to a reference model with $c^2_{\rm eff}=c^2_{\rm vis}=1/3$. Solid red lines and dashed red lines correspond to a $c^2_{\rm eff}$ of $0.36$ and $0.30$ respectively, whereas solid blue lines and dashed blue lines correspond to a $c^2_{\rm vis}$ of $0.36$ and $0.30$ respectively. Top and bottom panels are almost identical, showing that the relative effect of $(\ceffs$, $\cviss)$ is independent of neutrino masses.}\label{FIG:Cl}
\end{figure}

In figure~\ref{FIG:Cl} we show the CMB power spectra of our four models with non-standard values of $c^2_{\rm eff}$ and $c^2_{\rm vis}$, normalised to the reference  model with standard neutrino properties. The left column shows the ratio of the temperature power spectra, whereas the right column shows the ratio of the $E$-mode power spectra.

The CMB is sensitive to neutrino perturbations through gravitational interactions~\cite{Bashinsky:2003tk,Hou:2011ec,Lesgourgues:1519137}. The amplitude of photon oscillations (i.e., acoustic waves) is usually boosted near the time of SH crossing by the decay of metric fluctuations. In the  presence of a smooth free-streaming component like standard neutrinos, metric fluctuations get extra damping and the boosting is enhanced. After that time, photon perturbations oscillate with a higher amplitude on sub-SH scales. The enhanced boosting also implies that the phase of oscillations in the photon-baryon fluid is slightly shifted towards earlier times in presence of neutrinos. In the observable temperature and polarisation spectra, this induces a small displacement of CMB peaks towards larger angular scales. This ``neutrino drag'' effect is very characteristic of the presence of relativistic particles in the universe before photon decoupling~\cite{Bashinsky:2003tk}.\footnote{Instead of probing the existence of the cosmic neutrino background by varying the effective parameters $\ceffs$ and  $\cviss$ , one could directly introduce a parametrization of the CMB phase and investigate observational constraints on this phase, see\cite{follin}.} 

In the temperature power spectrum, the most prominent effect of $\ceffs$ and  $\cviss$ is a change in the amplitude of the spectrum, caused by different amounts of gravitational boosting. A lower $\ceffs$ leads to more density contrast in the neutrino species, so the metric fluctuations decay more slowly near SH crossing. The boosting of photon perturbations is reduced and the amplitude of the CMB  fluctuations is lower. The effect of $\cviss$  is  less straightforward,
since it impacts the evolution of $\Psi$ and $\Phi$ in a different way around the time of SH crossing for neutrino perturbations, i.e., near the time at which the neutrino anisotropic stress grows more or less fast, reaches a maximum and decays. For a smaller $\cviss$, the neutrino anisotropic stress is smaller at the time when the gravitational boosting of photon fluctuations is relevant, and this results in larger fluctuations. The change of amplitude observed in figure~\ref{FIG:Cl} is qualitatively different in the case of $\ceffs$ and $\cviss$, and is also different from a change in the primordial amplitude $A_s$, since it does not affect  scales that are above the SH at decoupling: it reaches a constant amplitude only for multipoles with roughly $\ell>300$, thus affecting the first and the second peak of the CMB in different ways and thereby changing  the shape of the spectrum.

Besides the oscillation amplitude, the parameters ($\ceffs$, $\cviss$) also change the phase of the acoustic oscillations, as one can see from the oscillatory patterns in figure~\ref{FIG:Cl}. Indeed we have seen in the previous section that the oscillation period of $\delta_\nu$ depends slightly on ($\ceffs$, $\cviss$). This shift is propagated to the photon-baryon fluid through the neutrino drag effect. In the polarisation power spectrum we find effects similar to those present in the temperature power spectrum. However, although the change in amplitude is similar to the one in the temperature power spectrum, the shift in the position of the peaks is even more clear, because for polarisation there is no contribution from Doppler effects.This explains the strong oscillations in the ratios shown in the right column of figure~\ref{FIG:Cl}.

By comparing the top and bottom panel of figure~\ref{FIG:Cl}, we see that the relative effect of $(\ceffs$, $\cviss)$ does not seem to depend on mass, even though the underlying power spectra do depend on mass. This is not unexpected.  When neutrinos have a small mass and become non-relativistic after photon decoupling, they affect the CMB through small effects: shift in the diameter angular distance, early integrated Sachs-Wolfe effect, and weak lensing. The first effect is totally independent of perturbations, and hence of $(\ceffs$, $\cviss)$. The second and third effects can in principle be affected by $(\ceffs$, $\cviss)$, but since this is a modulation of a small effect by another small effect, the impact of the effective speeds and of neutrino masses are independent of each other to a very good approximation. Hence the effect of neutrino masses cancels out in the $C_l$ ratios shown in figure~\ref{FIG:Cl}, at least in the neutrino mass  range explored here.

\subsection{Matter power spectrum}

 We complete  the previous analysis of the effects on the CMB power spectra of the effective parameters  ($\ceffs$, $\cviss$) with an analysis of potential signatures on the large scale structure of the universe, focusing on  the  shape of the matter power spectrum at redshift $z=0$.

\begin{figure}
\begin{center}
\includegraphics[width=0.45\textwidth]{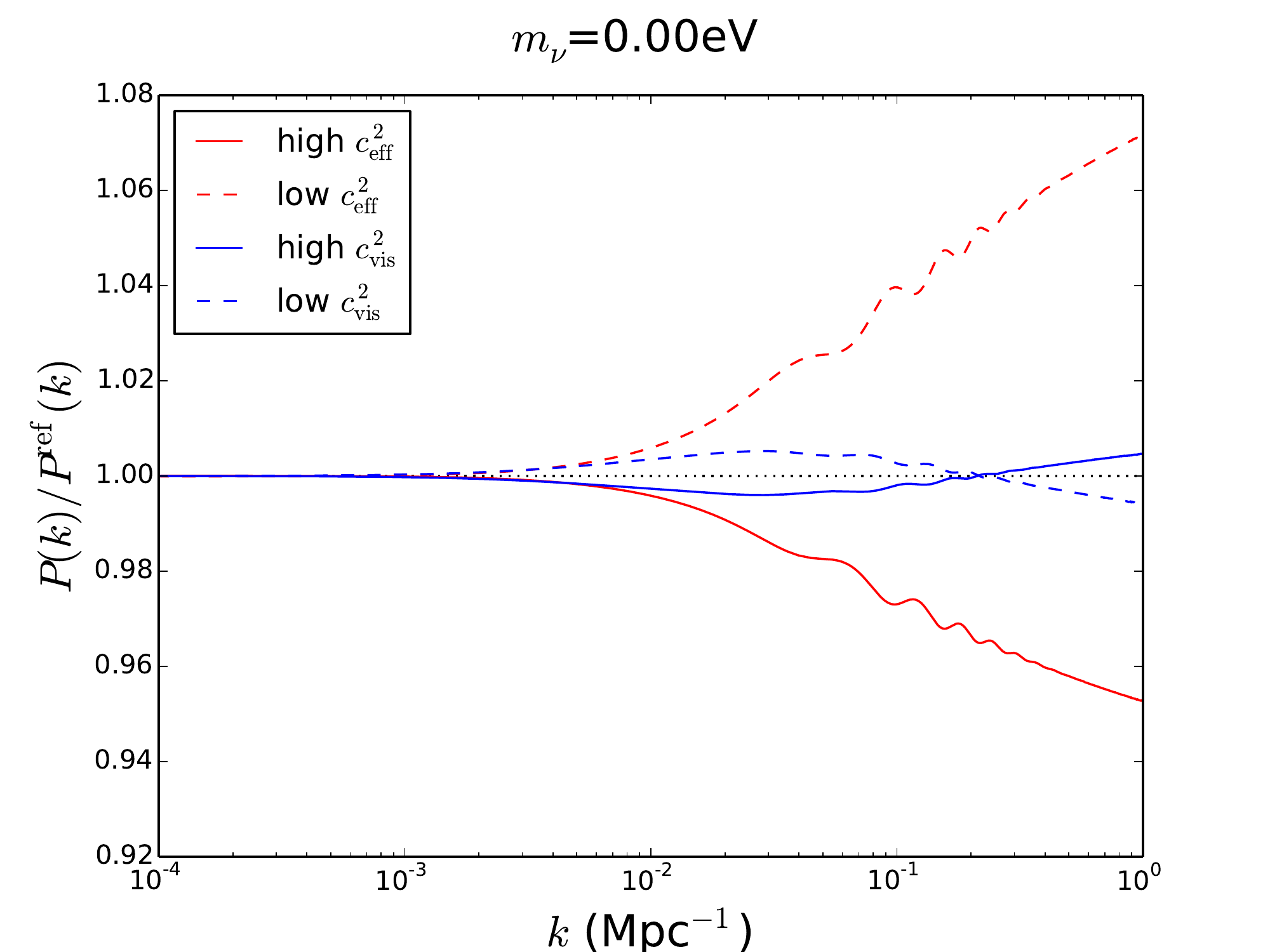}
\includegraphics[width=0.45\textwidth]{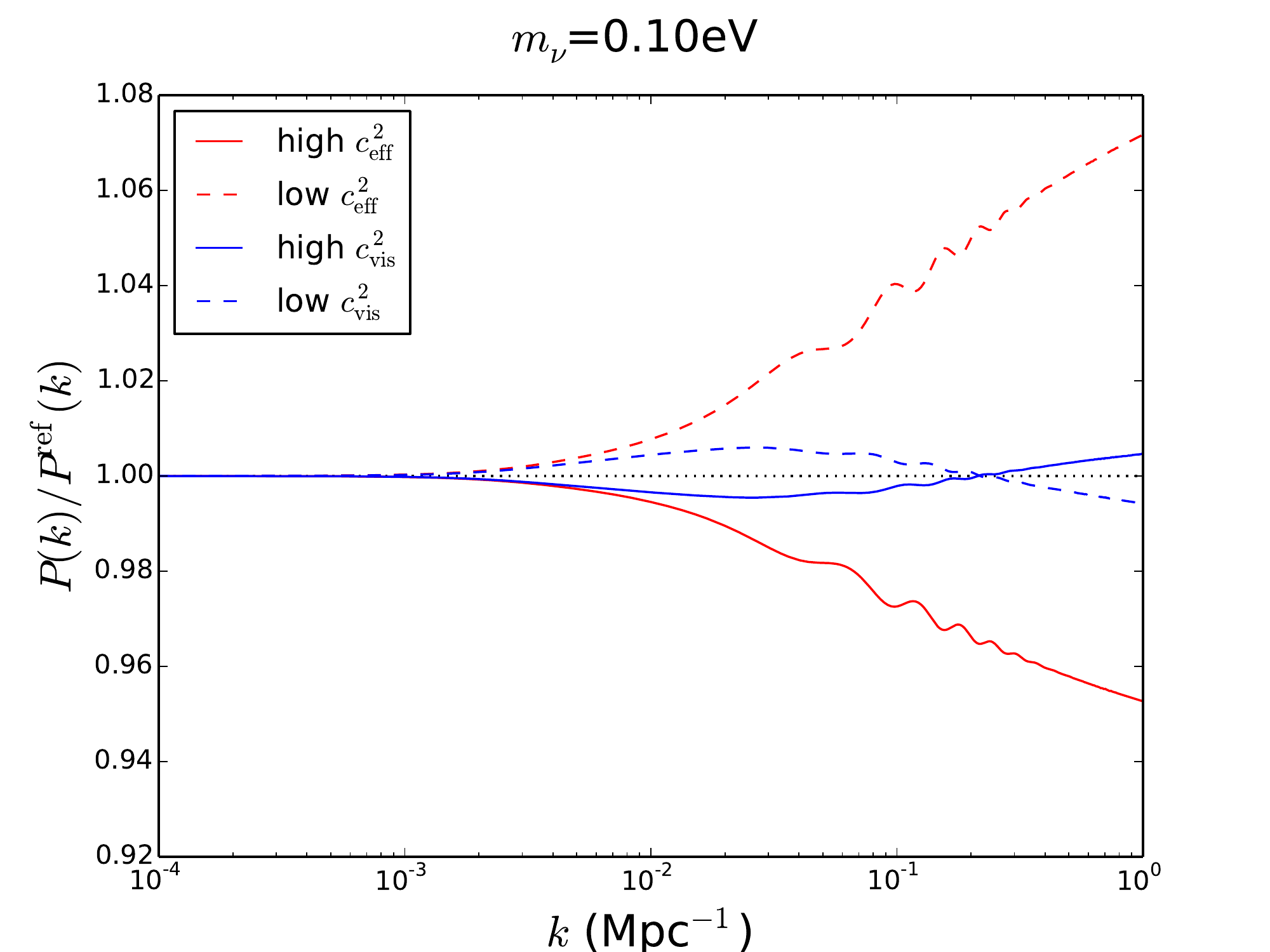}
\end{center}
\caption{Matter power spectrum for a $\Lambda$CDM model with massless neutrinos (left panel) and three degenerate neutrinos with $m_{\nu}=0.10$~eV each (right panel). All models are normalised to a reference model with $c^2_{\rm eff}=c^2_{\rm vis}=1/3$. Solid red lines and dashed red lines correspond to $c^2_{\rm eff}=0.36$ and $0.30$ respectively, whereas solid blue lines and dashed blue lines correspond to $c^2_{\rm vis}=0.36$ and $0.30$ respectively. These two plots are almost identical, showing that the relative effect of $(\ceffs$, $\cviss)$ is independent of neutrino masses.}\label{FIG:pk}
\end{figure}

In figure~\ref{FIG:pk}, as in the previous subsection, we plot the ratios of our four non-standard models with respect to the reference model with standard neutrinos.
On large scales ($k \lesssim 10^{-2}$Mpc$^{-1}$) the effects of these non-standard values of $c^2_{\rm eff}$ and $c^2_{\rm vis}$ are below 1\%. However, on smaller scales the effects become more important, especially for $c^2_{\rm eff}$.

The effect of $c^2_{\rm eff}$ on the matter power spectrum is easy to understand. Once the neutrino or dark radiation particles are non-relativistic, they fall into the gravitational potential wells of Cold Dark Matter. The growth rate of $\delta_\nu$ is larger than the one of $\delta_\mathrm{cdm}$ until the neutrino overdensities matches the CDM overdensities. We have seen in section 3.1. that for a smaller $c^2_{\rm eff}$, the density contrast $|\delta_\nu|$ starts growing a bit earlier and  from a slightly larger equilibrium value. Hence, the ratio $\delta_\nu/\delta_\mathrm{CDM}$ at a given scale and given time is larger for smaller $c^2_{\rm eff}$.

The growth rate of CDM and baryon fluctuations is slightly reduced when neutrino perturbations are negligible. With  a smaller $c^2_{\rm eff}$, there is a larger density contrast $|\delta_\nu|$ in the neutrinos, hence CDM and baryon collapse at a slightly faster rate and the small-scale matter power spectrum is enhanced.

At scales between 0.01 and 0.2 Mpc$^{-1}$  increasing (decreasing) any of the two sound speed parameters cause a decrease (increase) in the power spectrum. This amplitude modulation is still below 1\% when we change $c^2_{\rm vis}$ within the limits explored here, but $c^2_{\rm eff}$ can introduce modulations of up to 5\% within the range 0.30-0.36.
Interestingly, at $k=0.2$ Mpc$^{-1}$ the modulation due to $c^2_{\rm vis}$ changes its sign and an increase in its value produces a decrease of the power spectrum, however the effect  remains  below 1\% even at $k=1$ Mpc$^{-1}$ for the range considered here.  As in the CMB power spectrum, we also detect no relative effects of the neutrino mass on these ratios.
These considerations indicate that the effect of a modest change in $c^2_{\rm eff}$ is relatively large in the shape of the matter power spectrum: large volume, forthcoming large-scale structure surveys should have the statistical power to measure sub-percent effects on these scales. For these reasons, it would be interesting to compare our results with those of \cite{Palanque-Delabrouille:2013gaa,Palanque-Delabrouille:2014jca}, where the authors use Lyman-$\alpha$ forest data to get constraints on massive neutrinos. Moreover the different behaviour of the two parameters on scales $k \gtrsim 0.1$~Mpc$^{-1}$ means that  any  degeneracy  between the two parameters can be lifted. 

\newcommand{\cvissq}{c^2_{\textrm{vis}}}
\newcommand{\ceffsq}{c^2_{\textrm{eff}}}
\section{Models and data set}

We consider six different models. All models share  the six parameters of the flat $\Lambda$CDM model,
with the additional $\cvissq$ and $\ceffsq$:
\begin{align*}
\{\omega_b, \omega_{\textrm{cdm}}, h, A_s, n_s, \tau_{\textrm{reio}},
\cvissq, \ceffsq\}.
\end{align*}
The first six cosmological parameters denote the baryon and
cold dark matter physical densities, the reduced Hubble parameter, the
amplitude and tilt of the initial curvature power spectrum at the pivot scale
$k_* = 0.05/$Mpc, and the optical depth to reionisation. The effective parameters $\cvissq$ and
$\ceffsq$ have been described in section~\ref{sec:model}.

\subsection{Model descriptions}
Since this 8-parameter model is our ``minimal'' model, we refer to it as ``M''.
We further explore possible degeneracies between $(\cvissq, \ceffsq)$ and the
total neutrino mass $M_\nu \equiv \sum m_{\nu}$ and/or the effective number of relativistic species
$N_{\textrm{eff}}$. These 3 additional models are referred to as M$+m_{\nu}$, M$+N_{\textrm{eff}}$ and M$+m_{\nu}{+}N_{\textrm{eff}}$ and have 9, 9 and 10 parameters  respectively.

We also check for degeneracies with  the dark
energy equation of state parameter $w$ (and this model is referred to as  M$+w$), and the running of the
primordial spectrum tilt $\alpha_s \equiv {dn_s}/{d\log k}$ (model called M$+\alpha$).

Unless otherwise stated, when parameters take a fixed value  we adopt the same settings as in the ``base model'' of the Planck 2013 parameter paper~\cite{PlanckXVI}.
In particular, when the neutrino mass is not a free parameter,
we assume two massless neutrino species, and one species with a small mass $m_\nu=0.06$~eV, motivated by the minimal values in the normal hierarchy scenario. In that case, we
assign the same $(\cvissq, \ceffsq)$ to the massless and massive species. We checked explicitly that the bounds on $(\cvissq, \ceffsq)$ obtained in that way are indistinguishable from what we would get by assuming three massless families. Indeed, a neutrino mass $m_\nu=0.06$~eV is too small to change the evolution of perturbations at CMB times, independently of the value of $(\cvissq, \ceffsq)$. Such a small mass affects the CMB only through a modification of the angular diameter distance to decoupling. Hence, like in the standard case with $(\cvissq, \ceffsq)$=$(1/3,1/3)$, the only impact of this fixed mass is a small shifting of the best-fit value of $H_0$ by roughly $-0.6$~km s$^{-1}$ Mpc$^{-1}$~\cite{PlanckXVI}.

When the neutrino mass is considered as a free parameter, we consider for simplicity three degenerate neutrinos with equal mass and $(\cvissq, \ceffsq)$ parameters, and the bounds we report are always on the 
total neutrino mass $M_\nu$. It is well-known that for a fixed total mass, current observations are not sensitive to the mass splitting between the three families of active neutrinos.

When $N_\mathrm{eff}$ is left free, we assume one massive species with $m_\nu=0.06$~eV and $N_\mathrm{ur}=N_\mathrm{eff}-1$ massless species, all with the same $(\cvissq, \ceffsq)$ (here `ur' stands for ultra-relativistic). Finally, when varying $N_\mathrm{eff}$ and $m_\nu$ at the same time, we take one massive species with mass $m_\nu$, and $N_\mathrm{ur}=N_\mathrm{eff}-1$ massless ones, all with the same $(\cvissq, \ceffsq)$. Of course, the decision to assign the same $(\cvissq, \ceffsq)$ to all species in all cases is somewhat arbitrary. For instance, it could be the case that only one species of neutrinos has significant interactions with a dark sector. This choice is dictated by simplicity. Also, as long as everything keeps being consistent with standard neutrino perturbations, this choice will probably be sufficient in order to establish whether more complicated models are worth investigating.

\subsection{Data sets and parameter extraction}
The parameter extraction is done with the public code {\sc Monte
Python}~\cite{Audren:2012wb}, using the Metropolis Hastings algorithm, and a
Cholesky decomposition in order to better handle the large number of nuisance
parameters~\cite{Lewis:2013hha}.
We adopt flat priors on all cosmological parameters. 
We also use importance sampling for exploring
small deviations to the posterior coming from additional datasets. We compare our six different models to 3 sets of experiments. 

The CMB set includes the Planck \cite{Planckexp} temperature power spectrum \cite{PlanckCl},  the low-$\ell$ information from {\sc WMAP}
polarisation~\cite{Bennett:2012zja}, as well as high-$\ell$ {\sc
ACT}~\cite{Sievers:2013ica} and {\sc SPT}~\cite{Reichardt:2011yv} data~\cite{PlanckXVI}.
The adopted Planck  likelihood functions are  the low-$\ell$ Commander
likelihood and  the high-$\ell$ CAMspec \cite{PlanckCl}.
The  CMB+lensing set contains in addition
the Planck lensing reconstruction \cite{PlanckLensing}.
The recent expansion history of the Universe as measured via the   Baryon Acoustic Oscillations (BAO)  technique is also considered as an additional data set  and  we use the determinations of refs. ~\cite{Anderson:2013zyy, Beutler:2011hx, Font-Ribera:2013wce, Ross:2014qpa}.

\section{Results}

{\bf $\Lambda$CDM+$c_{\rm eff}^2$+$c_{\rm vis}^2\equiv$ M}: Results for the  minimal model $\Lambda$CDM+$c_{\rm eff}^2$+$c_{\rm vis}^2$ (M) are reported in tables~\ref{tab:CMB},~\ref{tab:CMBlensing},~\ref{tab:CMBlensingBAO} for the three
different datasets, and illustrated by the left panel of
figure~\ref{results_1}. The standard values ($c_{\rm eff}^2$, $c_{\rm vis}^2$) are always well within the 95\% confidence intervals, so the data gives no indication of exotic physics in the dark radiation sector. These findings can be seen as further evidence in favour of the detection of  the cosmic neutrino background. Our results in this case reproduce those of ref.~\cite{Gerbinoetal} and confirm that current data are sensitive to $\cviss$ and especially to $c_{\rm eff}^2$. The effect of  the neutrino anisotropic stress is detected albeit at small statistical significance: $\cviss=0$ is disfavoured  at the 2.5$\sigma$ level  for CMB and CMB +lensing but (slightly) above   3$\sigma$ when BAO data are included. 
For all dataset combinations, we observe (figure~\ref{results_1}) a small anti-correlation between the two effective parameters. Indeed we have seen in section~3 that they affect the amplitude of CMB oscillations in different directions. Apart from the overall amplitude, their effects are clearly distinct as shown by figure~2 which explains the weakness of the correlation.

\begin{figure}[h!]
\includegraphics[width=\textwidth]{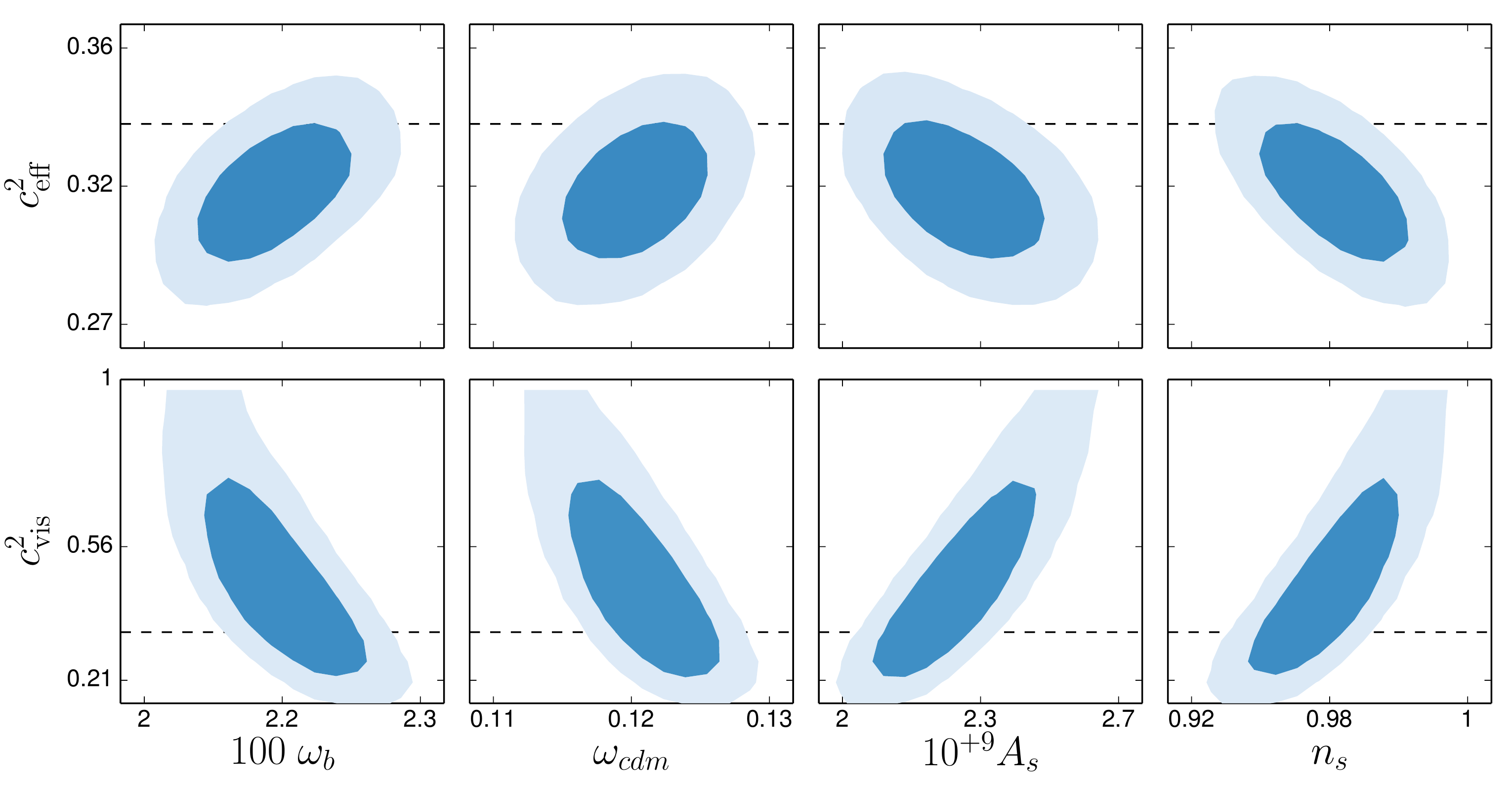}
\caption{Degeneracies between the parameters $(c^2_{\rm vis}, c^2_{\rm eff})$ and the parameters $\omega_b$, $\omega_{cdm}$, $A_s$ and $n_s$. A combination of CMB+lensing data is used for this plot, in which a $\Lambda$CDM+$c^2_{\rm vis}{+}c^2_{\rm eff}$ model is assumed. Dashed lines correspond to the standard values $(c_{\rm eff}^2, c_{\rm vis}^2)=(1/3, 1/3)$.}
\label{fig:degeneracies}
\end{figure}

\begin{figure}[h!]
\includegraphics[scale=0.4]{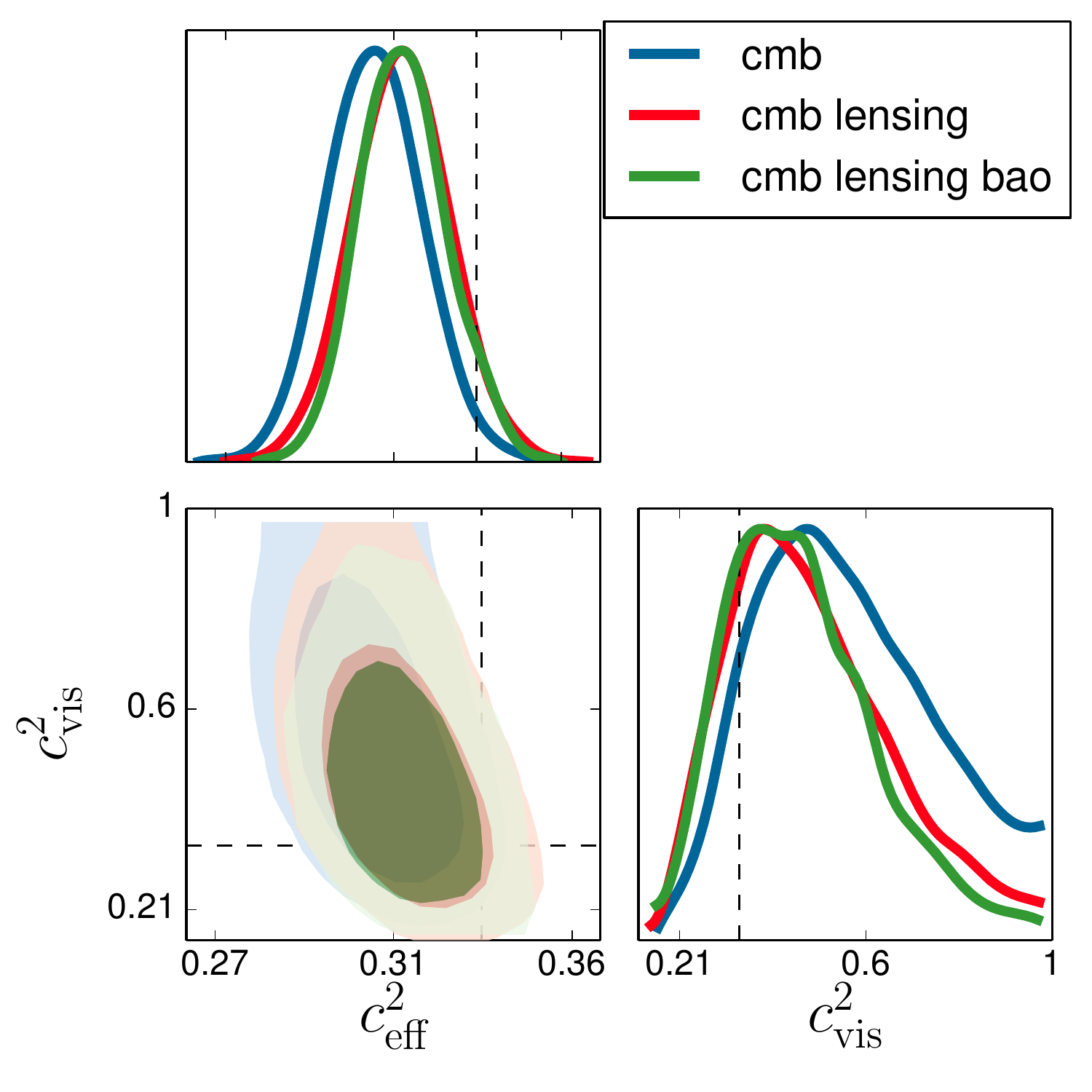}
\quad\quad\quad
\includegraphics[scale=0.32]{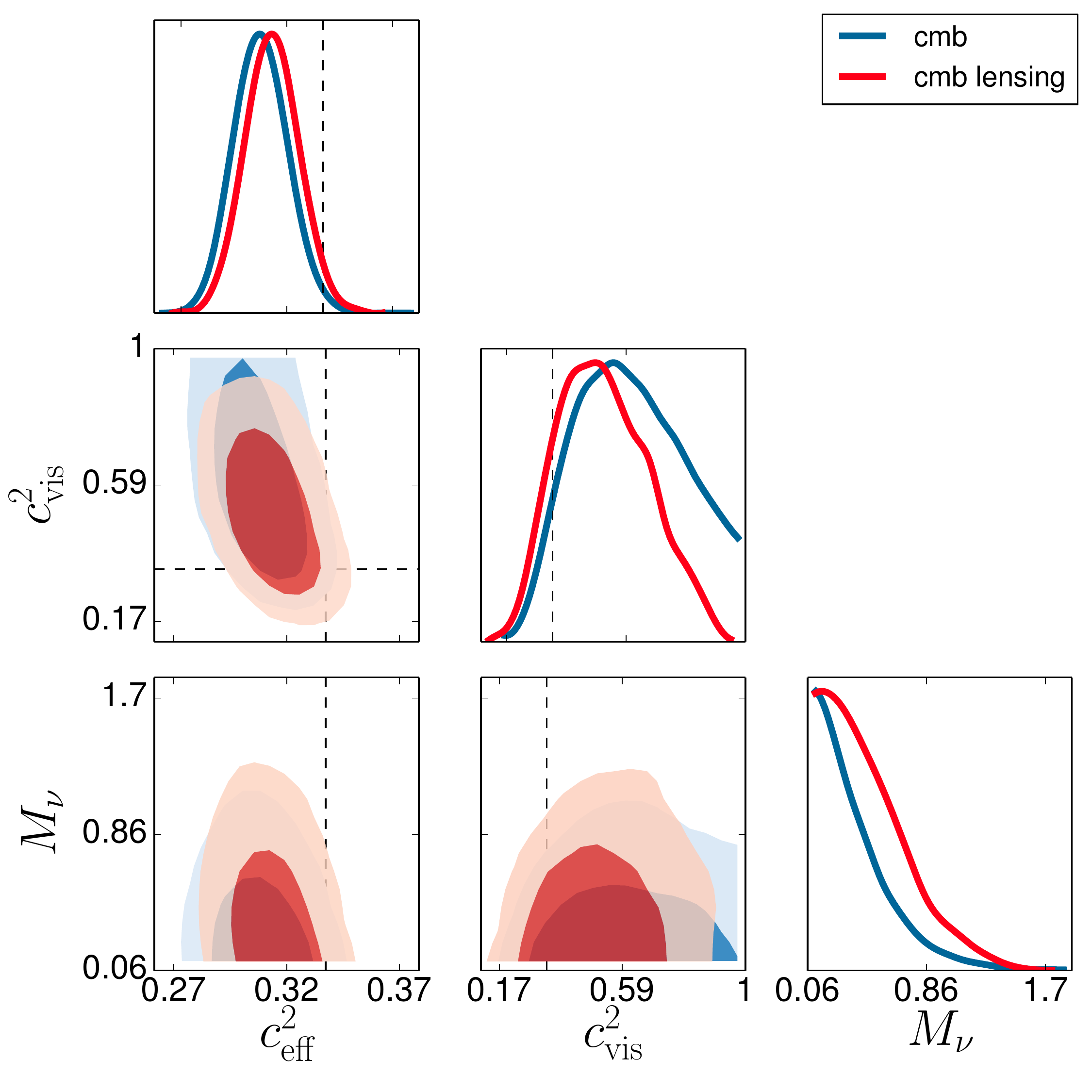}
\caption{\textit{Left}. 
Constraints in the ($c^2_{\rm vis}, c^2_{\rm eff}$) plane for combination of CMB, CMB+lensing and  CMB+lensing+BAO data, in the 
$\Lambda$CDM+ $c^2_{\rm vis}+c^2_{\rm eff}$ model. Marginalised posterior  distributions for both parameters are also shown.
\textit{Right.} 
Constraints on $(c^2_{\rm vis}$, $c^2_{\rm eff})$ and the total neutrino mass $M_\nu$ 
for CMB and CMB+lensing datasets in the $\Lambda$CDM+$c^2_{\rm vis}{+}c^2_{\rm eff}{+}m_{\nu}$ model. Dashed lines correspond to the standard values $(c_{\rm eff}^2, c_{\rm vis}^2)=(1/3, 1/3)$.}
\label{results_1}
\end{figure}

The bounds on the parameters of the $\Lambda$CDM model are significantly broader than in the
base $\Lambda$CDM case. In fact, the effect of $c_{\rm eff}^2$+$c_{\rm vis}^2$ discussed in section~3 turn out to be degenerate with subtle combinations of $\omega_b$, $\omega_{cdm}$, $n_s$ and $A_s$ (see figure~\ref{fig:degeneracies}). In particular, a high $\cviss$ requires low $\omega_b$, $\omega_{cdm}$, and high $n_s$ and $A_s$. Better CMB data could help break these degeneracies, and bring stronger constraints on ($c_{\rm eff}^2$, $c_{\rm vis}^2$).

This also indicates that the significance of the deviation from a scale invariant power spectrum relies on assuming standard neutrino properties. If this assumption is relaxed our knowledge of the shape of the primordial power spectrum is also degraded.

\begin{table}[ht]
\begin{center}
\begin{tabular}{|c||c|c|}
\hline
\multicolumn{3}{ |c| }{\textbf{CMB}} \\
\hline\hline  \rule{0pt}{3ex}
Parameter & $\Lambda$CDM+$c_{\rm eff}^2$+$c_{\rm vis}^2$ & + $m_\nu$ \\\hline  \rule{0pt}{3ex}
$100~\omega_{b }$ & $2.132_{-0.054}^{+0.044}$ & $2.107_{-0.056}^{+0.046}$ \\
$\omega_{cdm }$ & $0.1164 \pm 0.0040$ & $0.1166_{-0.0041}^{+0.0039}$ \\
$H_0$ & $68.0 \pm 1.3$ & $65.0_{-1.8}^{+3.4}$ \\
$10^{+9}A_{s }$ & $2.37 \pm 0.14$ & $2.40_{-0.13}^{+0.14}$ \\
$n_{s }$ & $0.991_{-0.019}^{+0.021}$ & $0.992_{-0.017}^{+0.022}$ \\
$\tau_{reio }$ & $0.090_{-0.014}^{+0.013}$ & $0.090_{-0.014}^{+0.013}$ \\
$c_{\rm eff}^2$ & $0.307_{-0.014}^{+0.013}$ & $0.304_{-0.014}^{+0.013}$ \\
$c_{\rm vis}^2$ & $0.56_{-0.25}^{+0.15}$ & $ 0.61_{-0.24}^{+0.17}$ \\
$M_\nu$~[eV] & -- & $< 0.88$ \\\hline
\end{tabular}
\end{center}
\caption{Constraints from CMB data on the values of the cosmological parameters for
the $\Lambda$CDM+$c_{\rm eff}^2$+$c_{\rm vis}^2$ and
the $\Lambda$CDM+$c_{\rm eff}^2$+$c_{\rm vis}^2+m_\nu$ models. 
We report the 95\%~C.L. upper limit for the total neutrino mass $M_\nu$, 
the mean values and $1\sigma$ ranges for all the other parameters.
}
\label{tab:CMB}
\end{table}

\begin{table}[ht]
\begin{center}
\begin{tabular}{|c||c|c|c|c|c|c|}
\hline
\multicolumn{7}{ |c| }{\textbf{CMB + lensing}} \\
\hline\hline \rule{0pt}{3ex}
Parameter & $\Lambda$CDM+$c_{\rm eff}^2$+$c_{\rm vis}^2$ & +$N_{\rm eff}$ & +$m_\nu$ & +$w$ & + $\alpha_s$ & + $N_{\rm eff} + m_\nu$\\ \hline \rule{0pt}{3ex}
$100~\omega_{b }$ & $2.162_{-0.052}^{+0.047}$ & $2.174_{-0.055}^{+0.057}$ & $2.124_{-0.056}^{+0.048}$ & $2.179_{-0.056}^{+0.052}$ & $2.180_{-0.056}^{+0.050}$ & $2.136_{-0.068}^{+0.060}$ \\
$\omega_{cdm }$ & $0.1163_{-0.0034}^{+0.0037}$ & $0.1181_{-0.0051}^{+0.0054}$ & $0.1186_{-0.0036}^{+0.0037}$ & $0.1164_{-0.0035}^{+0.0037}$ & $0.1163 \pm 0.0035$ & $0.1184\pm 0.0055 $ \\
$H_0$ & $68.3 \pm 1.1$ & $69.6 \pm 2.9 $ & $63.7_{-2.6}^{+4.1}$ & $85.5_{-4.5}^{+14.0}$ & $68.3_{-1.2}^{+1.1}$ & $65.4_{-4.2}^{+4.0}$ \\
$10^{+9}A_{s }$ & $2.31_{-0.15}^{+0.12}$ & $2.34_{-0.16}^{+0.12}$ & $2.36 \pm 0.13$ & $2.27_{-0.15}^{+0.12}$ & $2.35_{-0.15}^{+0.13}$ & $2.39\pm 0.14$ \\
$n_{s }$ & $0.984_{-0.020}^{+0.021}$ & $0.991_{-0.025}^{+0.024}$ & $0.981_{-0.018}^{+0.020}$ & $0.979_{-0.021}^{+0.022}$ & $0.980_{-0.019}^{+0.022}$ & $0.987_{-0.022}^{+0.025}$ \\
$\tau_{reio }$ & $0.090_{-0.014}^{+0.012}$ & $0.093_{-0.015}^{+0.013}$ & $0.093_{-0.014}^{+0.013}$  & $0.088_{-0.014}^{+0.012}$ & $0.095_{-0.016}^{+0.013}$ & $0.094_{-0.016}^{+0.013}$ \\
$c_{\rm eff}^2$ & $0.314 \pm 0.013$ & $0.314 \pm 0.013$ & $0.309_{-0.014}^{+0.013}$ & $0.318_{-0.014}^{+0.013}$ & $0.320_{-0.016}^{+0.014}$ & $0.312_{-0.013}^{+0.014}$ \\
$c_{\rm vis}^2$ & $0.49_{-0.22}^{+0.12}$ & $0.49_{-0.21}^{+0.11}$ & $0.51_{-0.19}^{+0.14}$ & $0.46_{-0.23}^{+0.11}$ & $0.50_{-0.22}^{+0.13}$ & $0.56_{-0.24}^{+0.14}$ \\
$N_{\rm eff}$ & -- & $3.22_{-0.37}^{+0.32}$ & -- & -- & -- & $3.17_{-0.37}^{+0.34}$\\
$M_\nu$~[eV] & -- & -- & $< 1.03$ & -- & -- & $< 1.05$  \\
$w$ & -- & -- & --  & $-1.49_{-0.38}^{+0.18}$ & -- & -- \\
$\alpha_{s }$ & -- & -- & -- & -- & $-0.010 \pm 0.010 $ & -- \\
\hline
\end{tabular} 
\end{center}
\caption{Constraints from CMB+lensing data on the values of the cosmological parameters for the
$\Lambda$CDM+$c_{\rm eff}^2$+$c_{\rm vis}^2$,
$\Lambda$CDM+$c_{\rm eff}^2$+$c_{\rm vis}^2+N_{\rm eff}$,
$\Lambda$CDM+$c_{\rm eff}^2$+$c_{\rm vis}^2+m_\nu$,
$\Lambda$CDM+$c_{\rm eff}^2$+$c_{\rm vis}^2+w$,
$\Lambda$CDM+$c_{\rm eff}^2$+$c_{\rm vis}^2+\alpha_s$
and $\Lambda$CDM+$c_{\rm eff}^2$+$c_{\rm vis}^2{+}N_{\rm eff}{+}m_\nu$ models. 
We report the 95\%~C.L. upper limit for the total neutrino mass $M_\nu$, 
the mean values and $1\sigma$ ranges for all the other parameters. }
\label{tab:CMBlensing}
\end{table}

\begin{table}[ht]
\begin{center}
\begin{tabular}{|c||c|c|}
\hline
\multicolumn{3}{ |c| }{\textbf{CMB + lensing + BAO}} \\
\hline\hline  \rule{0pt}{3ex}
Parameter & $\Lambda$CDM+$c_{\rm eff}^2$+$c_{\rm vis}^2$ & +$m_\nu$ \\
\hline  \rule{0pt}{3ex}
$100~\omega_{b }$ & $2.167_{-0.054}^{+0.048}$ & $2.145_{-0.058}^{+0.042}$ \\
$\omega_{cdm }$ & $0.1167_{-0.0023}^{+0.0020}$ & $0.1150_{-0.0025}^{+0.0023}$\\
$H_0$ & $68.25_{-0.65}^{+0.63}$ & $67.60_{-0.93}^{+0.98}$\\
$10^{+9}A_{s }$ & $2.30_{-0.12}^{+0.10}$ & $2.37\pm 0.13$\\
$n_{s }$ & $0.982_{-0.014}^{+0.017}$ & $0.992_{-0.014}^{+0.018}$ \\
$\tau_{reio }$ & $0.090 \pm 0.012$ & $0.094_{-0.014}^{+0.013}$\\
$c_{\rm eff}^2$ & $0.314_{-0.013}^{+0.011}$ & $0.309\pm 0.013$ \\
$c_{\rm vis}^2$ & $0.47_{-0.19}^{+0.12}$ & $0.54_{-0.18}^{+0.17}$ \\
$M_\nu$~[eV] & -- & $< 0.33$  \\
\hline
 \end{tabular}
\end{center}
\caption{Constraints from CMB+lensing+BAO data on the values of the
cosmological parameters for the
$\Lambda$CDM+$c_{\rm eff}^2$+$c_{\rm vis}^2$ and
$\Lambda$CDM+$c_{\rm eff}^2$+$c_{\rm vis}^2+m_\nu$ models.
We report the 95\%~C.L. upper limit for the total neutrino mass $M_\nu$,
the mean values and $1\sigma$ ranges for all the other parameters. }
\label{tab:CMBlensingBAO}
\end{table}

{\bf M+$m_\nu$}: The effect of adding $m_\nu$ can be seen in tables~\ref{tab:CMB},~\ref{tab:CMBlensing}, and in the right panel of figure~\ref{results_1}. There is no degeneracy between $c_{\rm eff}^2$+$c_{\rm vis}^2$ and the neutrino mass. This is an important new result, helping to establish the robustness of constraints on neutrino/dark radiation perturbations. Adding $m_\nu$ slightly decreases the mean value for $c^2_{\rm eff}$ and increases the mean value for $c^2_{\rm vis}$, but not by  a statistically significant amount.

\begin{figure}[h!]
\includegraphics[scale=0.32]{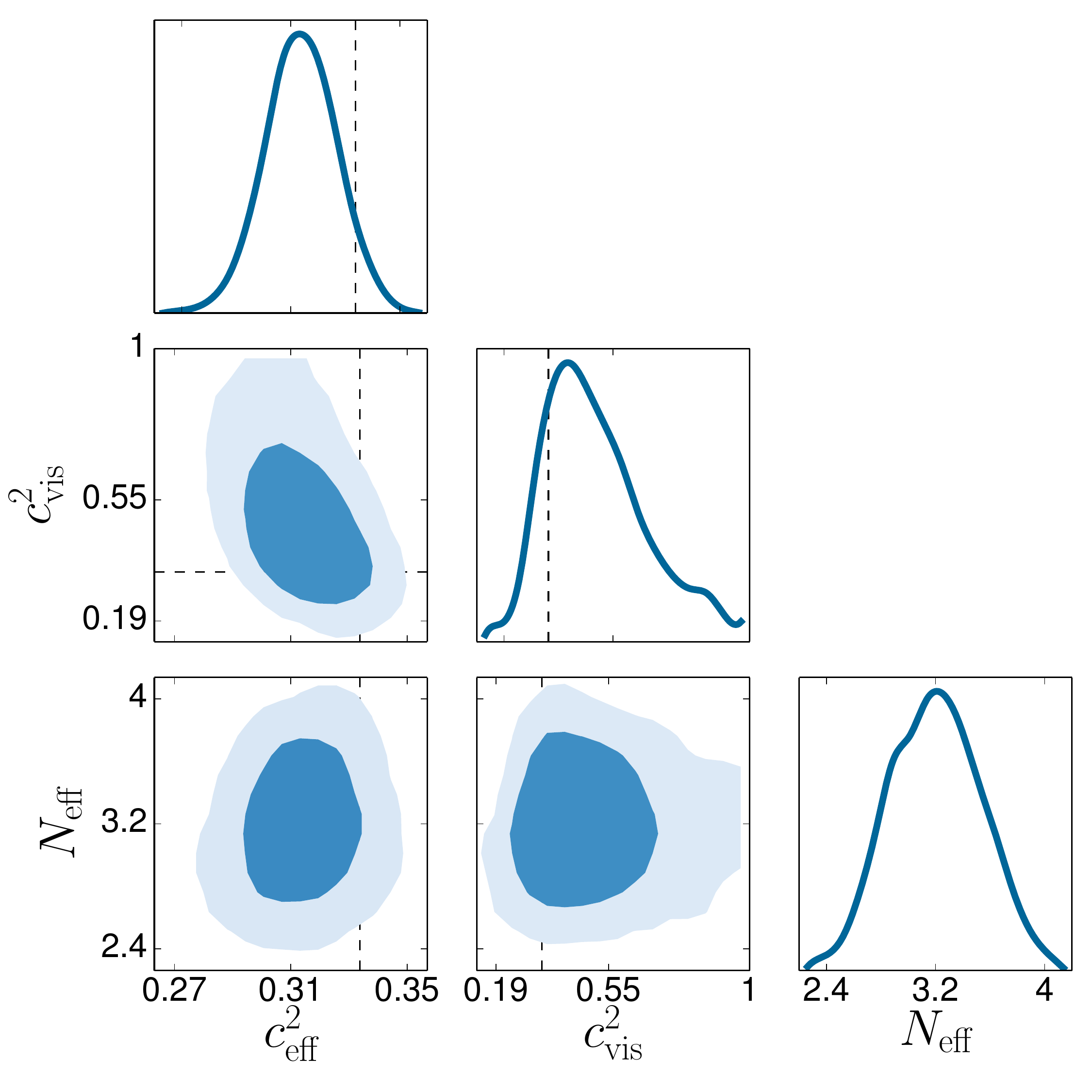}
\quad
\includegraphics[scale=0.32]{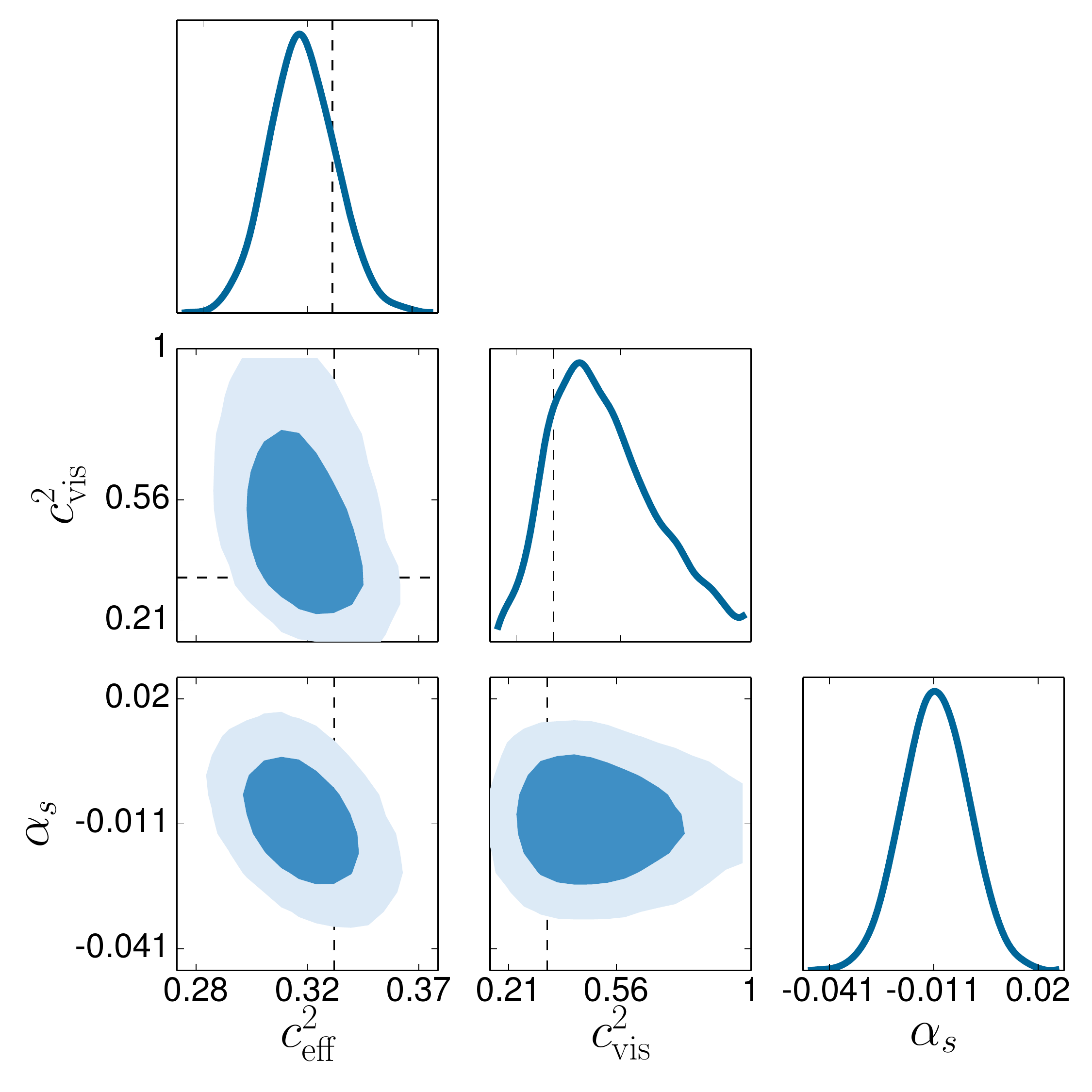}
\caption{\textit{Left.} Two-dimensional posterior distributions for 
$(c^2_{\rm vis}$, $c^2_{\rm eff})$ and $N_{\rm eff}$ for CMB+lensing data set in the 
$\Lambda$CDM+$c_{\rm eff}^2$+$c_{\rm vis}^2$+$N_{\rm eff}$ model, where we considered one massive ($m_\nu$=0.06~eV) and 
two massless neutrinos. 
\textit{Right.} Constraints on 
$(c^2_{\rm vis}$, $c^2_{\rm eff})$ and the running spectral index 
$\alpha_s$ for CMB+lensing data in the \mbox{$\Lambda$CDM+$c_{\rm eff}^2$+$c_{\rm vis}^2$+$\alpha_s$} model. Dashed lines correspond to the standard values $(c_{\rm eff}^2, c_{\rm vis}^2)=(1/3, 1/3)$.}
\label{results_2}
\end{figure}

\begin{figure}[h!]
\includegraphics[scale=0.25]{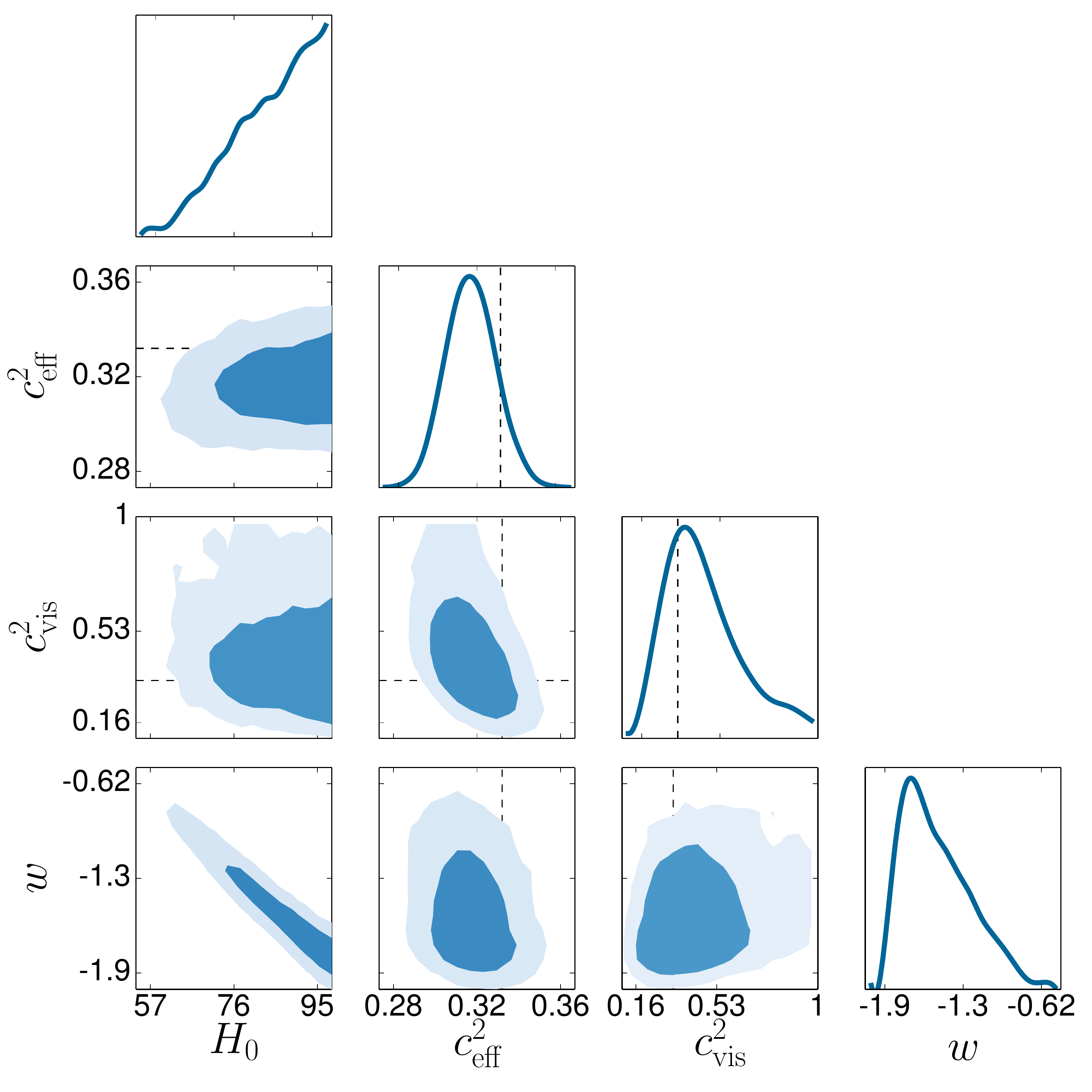}
\includegraphics[scale=0.25]{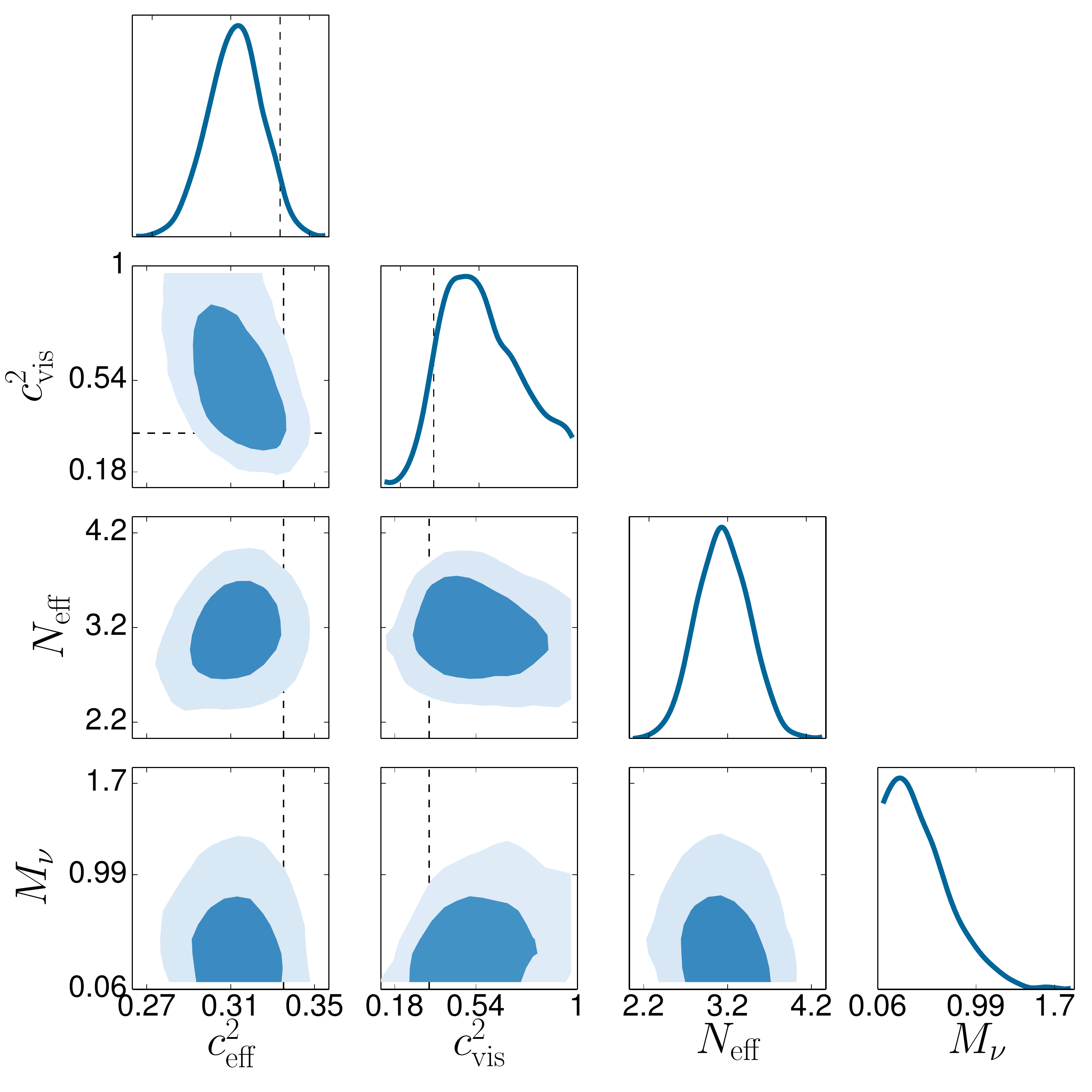}
\caption{\textit{Left.} Constraints on the interesting parameters for CMB+lensing data in the 
\mbox{$\Lambda$CDM $c_{\rm eff}^2$+$c_{\rm vis}^2$+$w$} model, where also the dark energy equation of state $w$ is let free to vary. 
We also include the contour in $H_0$ due to the strong degeneracy of this parameter with $w$. Remarkably, a prior on $H_0$ from direct measurements of the Hubble constant would break this degeneracy without changing our constraints on $(c^2_{\rm vis}$, $c^2_{\rm eff})$.
\textit{Right.} Constraints on the interesting parameters for CMB+lensing data in the $\Lambda$CDM+$c_{\rm eff}^2$+$c_{\rm vis}^2$+$N_{\rm eff}$+$m_{\nu}$ 
model. We report the constraints on the total neutrino mass $M_\nu$ in eV. Dashed lines correspond to the standard values $(c_{\rm eff}^2, c_{\rm vis}^2)=(1/3, 1/3)$. }
\label{results_3}
\end{figure}

{\bf Extended cosmologies}:
We considered extended cosmologies for the CMB+lensing dataset.
Parameter constraints are reported in table~\ref{tab:CMBlensing}. Selected two-dimensional posterior distributions  involving $(c_{\rm eff}^2$, $c_{\rm vis}^2)$ and the extra cosmological  parameters are shown in  figures~\ref{results_2} and~\ref{results_3}.
The  $(c_{\rm eff}^2$, $c_{\rm vis}^2)$  constraints are robust to the addition of extra cosmological parameters. There is no  significant degeneracy between  $(c_{\rm eff}^2$, $c_{\rm vis}^2$ and  $N_{\rm eff})$ or $w$. There is a small  anti-correlation between $c_{\rm eff}^2$ and $\alpha_s$ which however does not change the conclusion that $c_{\rm eff}^2$ is compatible with the standard value of 1/3  and  $\alpha_s$ is consistent with 0.

{\bf M+$N_{\rm rel}$+$m_{\nu}$}: Finally even  in the 10 parameters model where all parameters describing neutrino and dark radiation properties are left to vary we find no significant degeneracies with the  $c_{\rm eff}^2$, $c_{\rm vis}^2$ parameters. 
The effective number of species is still compatible with the standard value and its error-bar  ($\pm 0.34$) has  not  degraded compared to the $\Lambda$CDM+$N_{\rm eff}$ case ($\pm 0.33$) in \cite{PlanckXVI}. 
The 95\%  limit on the total neutrino mass is $M_{\nu}<1.05$ eV, which is only slightly degraded compared with the  $\Lambda$CDM+$m_{\nu}$ case $M_\nu < 0.85$ eV.

\section{Conclusions}

In this paper we have elucidated the physical effects of the $c^2_{\rm eff}$ and $c^2_{\rm vis}$ parameters on the CMB temperature and polarisation power spectra and the matter power spectrum. We find that the main signatures in the temperature and polarisation spectra are a shift of acoustic peaks, and a scale-dependent amplitude modulation for multipoles $\ell<300$ i.e., including the first peak, whereas the amplitude change is roughly constant beyond that scale and up to multipole $\ell=5000$. Interestingly, an increase in the $c^2_{\rm eff}$ parameter causes an increase in the amplitude, whereas an increase in the $c^2_{\rm vis}$ parameter causes the opposite effect. A similar amplitude change is found in the polarisation power spectrum. The matter power spectrum on the other hand, is mainly unaffected by these parameters at large scales, but it shows some dependence on these parameters at scales below matter-radiation equality. While $c^2_{\rm vis}$ effects are within 1\%, we find that $c^2_{\rm eff}$ can cause changes of several percent already at $k=0.2$~Mpc$^{-1}$ for the values we have studied.
Forthcoming large-scale structure surveys covering volumes of several Gpc$^3$  have in principle the statistical power to measure sub-percent effects on these scales. In practice, however, the  accurate determination of the shape of the matter power spectrum  and its interpretation in terms of the linear power spectrum on these scales is affected by other astrophysical  processes and it remains to be seen whether a sub-percent accuracy can be achieved realistically.

We have also investigated the existence of degeneracies between these dark energy perturbation effective parameters and cosmological parameters, such as the total neutrino mass $M_{\nu}$, effective number of relativistic species $N_{\rm eff}$, equation of state of dark energy $w$, and running of the spectral index $\alpha_s$. We note that our constraints on ($\omega_b$, $\omega_{\rm cdm}$, $A_s$, $n_s$) are significantly broader than in the standard case, but in this paper we concentrate on results for  $c^2_{\rm eff}$ and $c^2_{\rm vis}$ and on their degeneracies with extended cosmology parameters. We find that the $c^2_{\rm eff}$ and $c^2_{\rm vis}$ parameters are anti-correlated, that $\alpha_s$ is  slightly anti-correlated with $c^2_{\rm eff}$, but also that there are no major correlations between ($c^2_{\rm eff}$,~$c^2_{\rm vis}$) 
and $N_{\rm eff}$, and for the first time, we show that there is no significant correlation with the total neutrino mass $M_{\nu}$ either.

One can argue that our choice of constant $\ceffs$ and $\cviss$ is arbitrary and may not be sufficient to describe massive neutrinos from low momenta to high momenta. We have to bear in mind that these are effective parameters: in the absence of any significant deviations from their standard, constant,  values  they should be interpreted  in the light of a null test hypothesis. We can however go beyond this interpretation by assuming that $\ceffs$ depends on the momentum $q$ and expand this dependence to linear order: $c^2_{\rm eff}(q)=c^2_{{\rm eff}(0)}+c^2_{{\rm eff}(1)}\left(q-q_{\rm avg}\right)+\ldots$, where $q_{\rm avg}$ ($\simeq 3.15$) is the average momentum for neutrinos. From this expansion, it follows that being sensitive to $c^2_{{\rm eff}(1)}$ is equivalent to being sensitive to some $\ceffs$ for a relativistic momentum bin versus a non-relativistic momentum bin. On the other hand, a modification of the neutrinos mass produces a similar effect, since it regulates the time scale at which massive neutrinos become non-relativistic. In our analysis we found that, by fixing the values of $(\ceffs,\cviss)$, the dependence on the mass is negligible. This finding indicates therefore that our choice of constant $(\ceffs,\cviss)$ is a good approximation even for a $q$-dependent $\ceffs$.

Already with  state-of the art CMB data available (i.e., Planck 2013 data release and  WMAP low $\ell$ polarisation data)   alone  or in combination with other data sets (e.g., BAO), we can conclude that these parameters are not significantly degenerate with any other, and hence that the detection of the anisotropies of the cosmic neutrino background is robust.
We find no evidence for deviations from the standard neutrino model, i.e., 3 neutrino families with effective parameters ($c_{\rm eff}^2$, $c_{\rm vis}^2$)=(1/3,~1/3) when we consider CMB data only (including CMB lensing).

However the inclusion of  $(c_{\rm eff}^2$, $c_{\rm vis}^2)$ parameters degrades the constraints on some of the $\Lambda$CDM model parameters, such as the physical matter density and the slope of the primordial power spectrum.  In particular, high values of $n_s$, including a scale invariant power spectrum ($n_s=1$), become allowed.  This indicates that the significance of the deviation from a scale invariant power spectrum, with all its consequences for inflationary models,  relies on assuming standard neutrino properties. It also means that future data sets providing independent measurements of these parameters, such as the matter power spectrum from galaxy surveys or smaller scale CMB polarization, could help to remove degeneracies and greatly improve the sensitivity to $(c_{\rm eff}^2$, $c_{\rm vis}^2)$. This is expected to be the case for the full Planck data on temperature and polarisation anisotropies.  Measurements of the shape of the matter power spectrum,  even on linear scales, should also greatly help to lift the $\{n_s$, $c_{\rm eff}^2$, $c_{\rm vis}^2\}$ degeneracies.
\acknowledgments
This work results from a workshop at ICC (Barcelona) ``Tools for Cosmology" by JL, TT and BA 
(\url{http://icc.ub.edu/~liciaverde/ERCtraining.html}).
LV, AJC and EB are supported by the European Research Council under the European Community's Seventh Framework Programme FP7-IDEAS-Phys.LSS 240117. 
The work of EB is partially supported by "Fondazione Angelo Della Riccia".
LV acknowledges support by Mineco grant FPA2011-29678-C02-02.
VN acknowledges support by Spanish MINECO through project FPA2012-31880, by Spanish MINECO (Centro de excelencia Severo Ochoa Program) under grant SEV-2012-0249 and by the European Union through the FP7 Marie Curie Actions ITN INVISIBLES (PITN-GA-2011-289442).
VP acknowledges support by Labex grant ENIGMASS.
MPI acknowledges support from MINECO under the grant AYA2012-39702-C02-01.

Based on observations obtained with Planck (\url{http://www.esa.int/Planck}), an ESA science mission with instruments and contributions directly funded by ESA Member States, NASA, and Canada.

\bibliographystyle{utcaps}

\bibliography{nus}

\end{document}